\documentclass[aps,pre,twocolumn,longbibliography,superscriptaddress,amsmath,
amssymb,floatfix,nofootinbib,a4paper]{revtex4-2}

\usepackage{lineno}

\usepackage{latexsym}
\usepackage{wasysym}
\usepackage[english]{babel}
\usepackage{chngcntr}

\usepackage[makeroom]{cancel}

\usepackage{mathtools}
\usepackage{graphicx}
\usepackage[usenames,dvipsnames]{color}
\usepackage{xcolor}
\usepackage{epsfig}
\usepackage{wasysym}
\usepackage{natbib}
\usepackage{bm}
\usepackage[normalem]{ulem}

\newcommand*{\br}{{\bf r}}

\newcommand*{\bj}{{\bf j}}
\newcommand*{\be}{{\bf e}}
\newcommand*{\bff}{{\bf f}}
\newcommand*{\bn}{{\bf n}}

\newcommand*{\bg}{{\bf g}}
\newcommand*{\bv}{{\bf v}}
\newcommand*{\bs}{{\bf s}}
\newcommand*{\bw}{{\bf w}}

\newcommand*{\bu}{{\bf u}}
\newcommand*{\bF}{{\bf F}}

\newcommand*{\eq}[1]{Eq.~(\ref{#1})}
\newcommand*{\eqs}[1]{Eqs.~(\ref{#1})}

\newcommand*{\eref}[1]{(\ref{#1})}
\newcommand*{\sref}[1]{Sect.~\ref{#1}}
\newcommand*{\fref}[1]{Fig.~\ref{#1}}

\newcommand*{\Fref}[1]{Figure~\ref{#1}}
\newcommand*{\aref}[1]{Appendix~\ref{#1}}

\begin{document}

\title{Hydrodynamic Stokes flow induced by a chemically active patch imprinted 
on a planar wall 
}

\author{Mihail N. Popescu}
\email{\texttt{mpopescu@us.es}}
\affiliation{F\'\i sica Te\'orica, Universidad de Sevilla, Apdo.~1065, 
41080 Sevilla, Spain}

\author{Bogdan Adrian Nicola}
\email{\texttt{bnicola@biodyn.ro}}
\affiliation{Electrochemistry Laboratory, International Centre of Biodynamics, 1B Intrarea Portocalelor, 060101 Bucharest, Romania}
          
\author{William E. Uspal}
\email{\texttt{uspal@hawaii.edu}}
\affiliation{Department of Mechanical Engineering, University of Hawai'i at M{\=a}noa, 2540 Dole Street, Holmes Hall 302, Honolulu, HI 96822, USA} 
\affiliation{International Institute for Sustainability with Knotted Chiral Meta Matter (WPI-SKCM$^2$), Hiroshima University, 1-3-1 Kagamiyama, Higashi-Hiroshima, 739-8526 Hiroshima, Japan}

 \author{Alvaro Dom\'\i nguez}
\email{\texttt{dominguez@us.es}}
\affiliation{F\'\i sica Te\'orica, Universidad de Sevilla, Apdo.~1065, 
41080 Sevilla, Spain}
\affiliation{Instituto Carlos I de F{\'i}sica Te{\'o}rica y
  Computacional, 18071 Granada, Spain}         

\author{Szilveszter G\'asp\'ar}
\email{\texttt{sgaspar@biodyn.ro}}
\affiliation{Electrochemistry Laboratory, International Centre of Biodynamics, 1B Intrarea Portocalelor, 060101 Bucharest, Romania}

\begin{abstract}
Patches of catalyst imprinted on supporting walls induce motion of the fluid around them once they are supplied with the chemical species (``fuel'') that are converted by the catalytic chemical reaction. While the functioning of such chemically active micropumps is conceptually well understood, an in-depth characterization of the induced hydrodynamic flow, and in particular of its possible dependences on parameters such as material properties of the patch and the wall or the geometry of the experimental cell, remains elusive. By using a simple model for the chemical activity of a patch imprinted on a planar wall, we determine analytically the induced hydrodynamic flow in a Newtonian solution that occupies the half space above the wall supporting the patch. This can be seen as an approximation for an experimental-cell geometry with a height much larger than its diameter, the latter, in turn, being much larger than the size of the patch.) The general flow is a linear superposition of a surface-driven and a bulk-driven component; they have different topologies and, generically, each one dominates in distinct regions, with the surface-driven flow being most relevant at small heights above the wall. The surface-driven flows exhibit a somewhat unexpectedly rich behavior, including qualitative changes in the topology of the flow, as a function of the contrast in surface-chemistry (osmotic slip coefficient) between the patch and the support wall. The results are expected to provide guidance for the interpretation of the drift of tracers by the ambient flow, which is the observable usually studied in experimental investigations of chemically active micropumps.
\end{abstract}

\maketitle

\emergencystretch 3em

\section{Introduction}
\label{sec:Intro}

Microfluidic devices of varying complexity are currently utilized in areas ranging from fundamental research, such as the investigation of fluid transport problems, to the fabrication of commercially available laboratory equipment like genomic sequencing platforms \cite{nunes_introduction_2022,battat_outlook_2022}. Regardless of their area of application, these devices must be paired with a means for fluid manipulation. Fluid manipulation in microfluidic devices is most often performed using relatively bulky pumps that require an external power supply and rely on moving mechanical parts to generate negative or positive pressure drops. These pumps not only reduce the portability of the microfluidic devices but also add significant costs. As a result, substantial research effort is currently being directed towards developing novel methods for pumping and manipulation of fluids within microfluidic devices. 

For microfluidic channels, a promising alternative to traditional methods of driving fluid transport is the use of chemically active catalytic micropumps \cite{kline_catalytic_2005,zhou_chemistry_2016}. These are tiny structures, made of a 
catalyst and imprinted on planar walls, which, once supplied with the chemical species (``fuel'') involved in the catalytic chemical reaction that they promote, induce motion of the fluid solution around them by developing inhomogeneities in its chemical composition. One of the most important characteristics of catalytic micropumps is their ability to operate without an external power supply and without mechanically moving parts, which provides capabilities, to a certain degree, for autonomous operation (a useful feature, e.g., in a resource-poor environment).

Many aspects of catalytic micropumps have been revealed through the fabrication and study of inorganic catalytic pumps. Among the earliest such pumps were silver disks and rings, with diameters ranging from $\sim$ 6 to $\sim$ 120 $\mathrm{\mu m}$, microfabricated on gold surfaces \cite{kline_catalytic_2005,kline_catalytic_2006}; these were shown to stir aqueous solutions of hydrogen peroxide by the decomposition of hydrogen peroxide into water and oxygen. The induced flows were explained by electroosmosis occurring between the silver disks or rings (acting as a cathode) and the gold surface (acting as an anode). Similar observations of pumping by decomposition of hydrogen peroxide have been reported for platinum disks, with diameters of $\sim$ 30 to $\sim$ 50 $\mathrm{\mu m}$, microfabricated on gold surfaces \cite{farniya_imaging_2013}, and with silver disks of a diameter of 5 mm fabricated on glass surfaces \cite{gentile_silver-based_2020}. Pumping in the presence of hydrazine (instead of hydrogen peroxide) was achieved by replacing silver or platinum with palladium in the structure of these catalytic micropumps \cite{ibele_hydrazine_2007}. Micropumps made using semiconducting materials, such as titanium dioxide and doped silicon, which are  photocatalysts for water splitting and thus pump when illuminated at suitable light wavelength, have also been reported \cite{hong_light-driven_2010,yu_microchannels_2020,esplandiu_silicon-based_2015}. These photocatalytic pumps require no other fuel than water, and the light wavelength and its intensity can conveniently be used to turn these photocatalytic micropumps on and off and to control the amplitude of the flows produced by them \cite{yu_microchannels_2020}. 
Dissolution, depolymerization, and ion-exchange are a few other alternatives to catalytic processes when it comes to producing concentration gradients. Thus, not surprisingly, 
pumps based on the dissolution of microparticles \cite{mcdermott_self-generated_2012}, on the depolymerization of polymer patches \cite{zhang_self-powered_2012}, on the  release of a surfactant via a photoisomerisation process at an immobilized particle \cite{Svetlana2020,Svetlana2022}, and on ion-exchange resins \cite{niu_microfluidic_2017,esplandiu_radial_2022} have also been reported. Ref. \cite{esplandiu_radial_2022}, in particular, provides clear demonstrations that directional pumping over large distances is possible with surface active patches following the minimal three-materials strategy theoretically proposed by \cite{Michelin2019}.

The inorganic catalytic micropumps mentioned above are ingenious devices that meet the fluid manipulation requirements of some applications. However, their use in the biomedical field is limited due to their restricted biocompatibility; consequently, with biomedical applications in mind, enzyme micropumps have also been developed. Disk-shaped patches of various enzymes, including catalase \cite{sengupta_self-powered_2014,ortiz-rivera_enzyme_2016}, lipase \cite{sengupta_self-powered_2014}, urease \cite{sengupta_self-powered_2014,ortiz-rivera_convective_2016,gao_geometric_2022,alarcon-correa_self-assembled_2019}, glucose oxidase \cite{sengupta_self-powered_2014,munteanu_glucose_2019,munteanu_impact_2021}, DNA polymerase \cite{sengupta_dna_2014}, and acid phosphatase \cite{valdez_solutal_2017}, or even two enzymes ($\beta$-glucosidase and urease), which allows for control of the direction of pumping \cite{GNP_2024,Sen_dual_2024}, have been shown to pump/stir solutions containing the appropriate enzyme substrates. These pumps are not only better suited for biomedical applications but also significantly extend the range of fuels with which catalytic micropumps can function. Accordingly, by now there are various choices of active components available in order to adapt to specific requirements, e.g., biocompatibility. Moreover, such catalytic micropumps have been shown to be suitable, either individually or organized into arrays, for a variety of applications, in addition to pumping/stirring fluid in microfluidic devices. For example, they have been employed for controlled drug release \cite{sengupta_self-powered_2014}, the construction of colloidal crystals \cite{afshar_farniya_sequential_2014}, the directional delivery of microparticles \cite{das_harnessing_2017}, and sensing \cite{ortiz-rivera_enzyme_2016}. 

It is generally agreed that the flows induced by enzyme micropumps in the bulk solution seem to be primarily due to a solutal buoyancy effect, owing to the differences in the mass densities of the reactant and product molecules and the spatially inhomogeneous chemical composition induced by the activity at the patch
\cite{sengupta_self-powered_2014,valdez_solutal_2017,gentile_silver-based_2020}. On the other hand, 
in a number of systems, including the studies of silver patches in Ref. \cite{gentile_silver-based_2020}, of platinum disks in Ref. \cite{farniya_imaging_2013}, or of glucose oxidase (GOX) patches in Refs. \cite{munteanu_glucose_2019,munteanu_impact_2021}, it was observed that the motion of tracers in the vicinity of the wall on which the active patch is imprinted cannot be attributed solely to a bulk driven flow. These additional contributions have been heuristically attributed to both a phoretic response of the tracer to the chemical composition inhomogeneities and to a drift by flows produced via an osmotic-slip induced at the wall by the same chemical composition inhomogeneities \cite{farniya_imaging_2013,niu_microfluidic_2017,munteanu_glucose_2019}.  Moreover, in the \sref{sec:exper_motiv} below we discuss a new set of 
experiments in which it is observed that the direction of the tracer motion 
in a region located above the wall and outside, but near the edge, of the 
active patch can be reversed by changing the surface-chemistry of the wall, while keeping everything else the same. This strongly suggests that 
the tracer motion in this case is dominated by a drift due to osmotic flows induced at the wall by the chemical activity of the patch. Furthermore, contributions from flows driven by a self-induced electric field are to be expected, and to play a significant role, in the case when the catalytic reaction at the patch releases long-lived ionic species in the surrounding solution, e.g., in the systems studied by Refs. \cite{farniya_imaging_2013,esplandiu_radial_2022,niu_microfluidic_2017}.

Theoretical work on these types of flow has been, in general, connected with specific experimental realizations; due to the inherent complexities of the detailed modeling (e.g., dealing with a flow, driven by bulk force distributions, within a closed container \cite{valdez_solutal_2017}, or, as in the case where ionic species are involved in the activity, dealing with non-linearly coupled equations governing the transport of species and the electric potential \cite{farniya_imaging_2013,esplandiu_radial_2022}), in many cases only numerical studies have been possible \cite{sengupta_self-powered_2014,valdez_solutal_2017,das_harnessing_2017,ortiz-rivera_enzyme_2016,farniya_imaging_2013,esplandiu_radial_2022,ortiz-rivera_convective_2016,gao_geometric_2022,Michelin2015}. Analytical results for the flow profile within a microchannel with an active patch on a wall have been obtained for systems with strong confinement (quasi-two-dimensional geometries) \cite{gentile_silver-based_2020,niu_microfluidic_2017} or in planar slit geometry of arbitrary height but with a periodic distribution of patches \cite{davidson_predictive_2018,Michelin2019}.

From the discussion above, it is clear that part of the difficulties in achieving an in-depth understanding of the interplay between the bulk driven and surface driven flows, respectively, is due to the fact that the challenges raised by the complex nature of the physical problem (a non-equilibrium steady state of hydrodynamic flow induced by the chemical activity) are interwoven with additional mathematical complications raised by the set-up of the system (confining walls, patches of sizes that are similar to the lengthscales of the lateral and of the vertical confinement, etc). It is thus natural to consider the complementary approach of starting with simpler, yet experimentally relevant, geometries, for which analytical results can be derived, and then build additional constraints of spatial confinement upon the base provided by the results in the simpler setup. Accordingly, in this work we analyze the question of the hydrodynamic flow induced in liquid suspension occupying a half space above a planar wall by a chemically active patch at the wall. (This may also be useful and relevant in what concerns the chemotactic behavior of active Janus particles near walls with spatially localized sources of fuel recently approached by Ref. \cite{Mancuso_2024}.) For simple models of chemical activity, Newtonian liquid, and simple geometries of the patch, in this set-up it is possible to solve analytically\footnote{The final results are expressed, though, as integral representations or series that have to be numerically calculated.} the Stokes equations with the corresponding surface (``osmotic slip'') and bulk (``solutal buoyancy'') sources of flow.  

\section{Additional experimental motivation}
\label{sec:exper_motiv}

As it has been noted in the Introduction, although it is generally agreed that bulk-driven, buoyancy body-forces play an important role in the generation of flows by chemically active patches, there are experimental observations strongly suggesting the simultaneous action of osmotic, surface-driven flows. In short, it is now well documented that the behavior of tracers in the vicinity of the active patch wall could be different from the one far from the wall \cite{gentile_silver-based_2020,munteanu_glucose_2019,farniya_imaging_2013,esplandiu_silicon-based_2015}.
For example, for the bimetallic systems, like platinum patch on gold wall (or vice-versa), as well as for the platinum micropumps on doped silicon, the studies by Refs. \cite{farniya_imaging_2013,esplandiu_silicon-based_2015} report an apparent dependence of the tracers motion on the exact nature of the system; however, 
it has proven difficult to disentangle the observed tracer motion into all the various possible sources for it. A similar situation occurred in the case of the GOX patches studied by Ref. \cite{munteanu_glucose_2019}, where the motion of tracers near the surface involves a superposition of drift by ambient hydrodynamic flow and phoresis due to the chemical gradients induced by the oxidation  of glucose at the active patch. 
\begin{figure}[!b]
\includegraphics[width = \columnwidth]{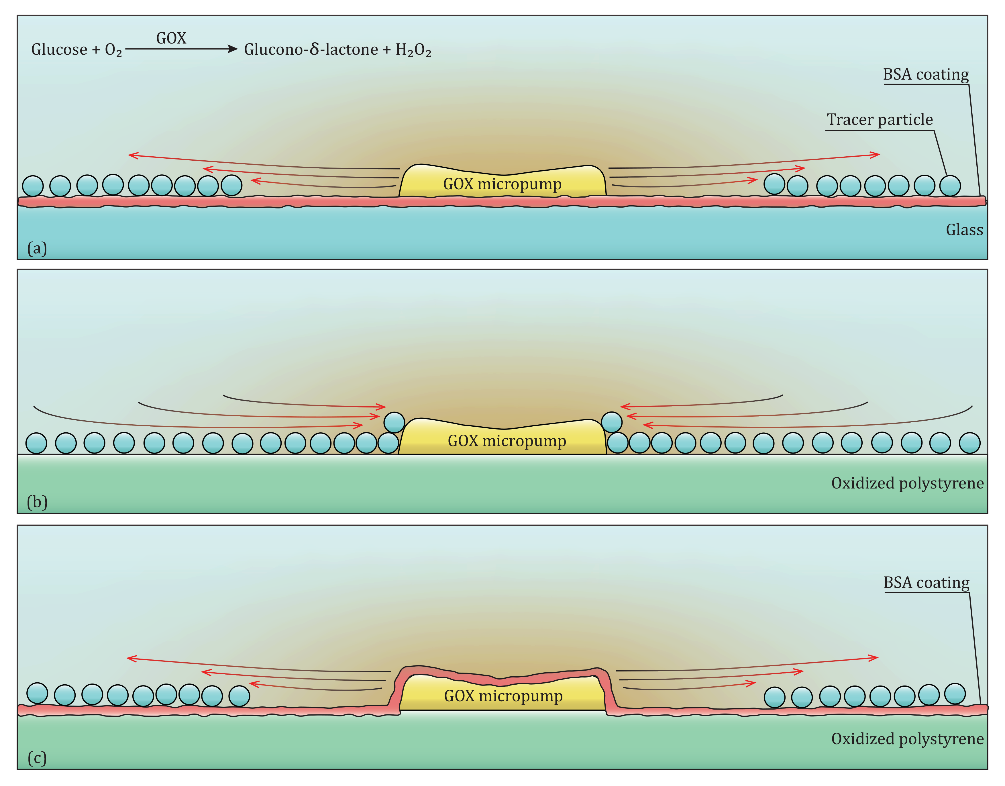}
\caption{\label{fig:schematic_exper} 
Schematic representation of a GOX patch imprinted onto (a) a BSA-coated glass surface, (b) an UV-oxidized polystyrene surface, and (c) an UV-oxidized polystyrene surface and then all coated with BSA. The biochemical reaction catalyzed by GOX is listed in the top left corner; the color gradient in the region above the patch depicts schematically a concentration gradient induced by the activity of the GOX patches, while the lines schematically show the observed motion of particles within the monolayer of sedimented tracers. 
}
\end{figure}

\begin{figure*}[!htb]
\begin{center}
\includegraphics[width=0.8\paperwidth]{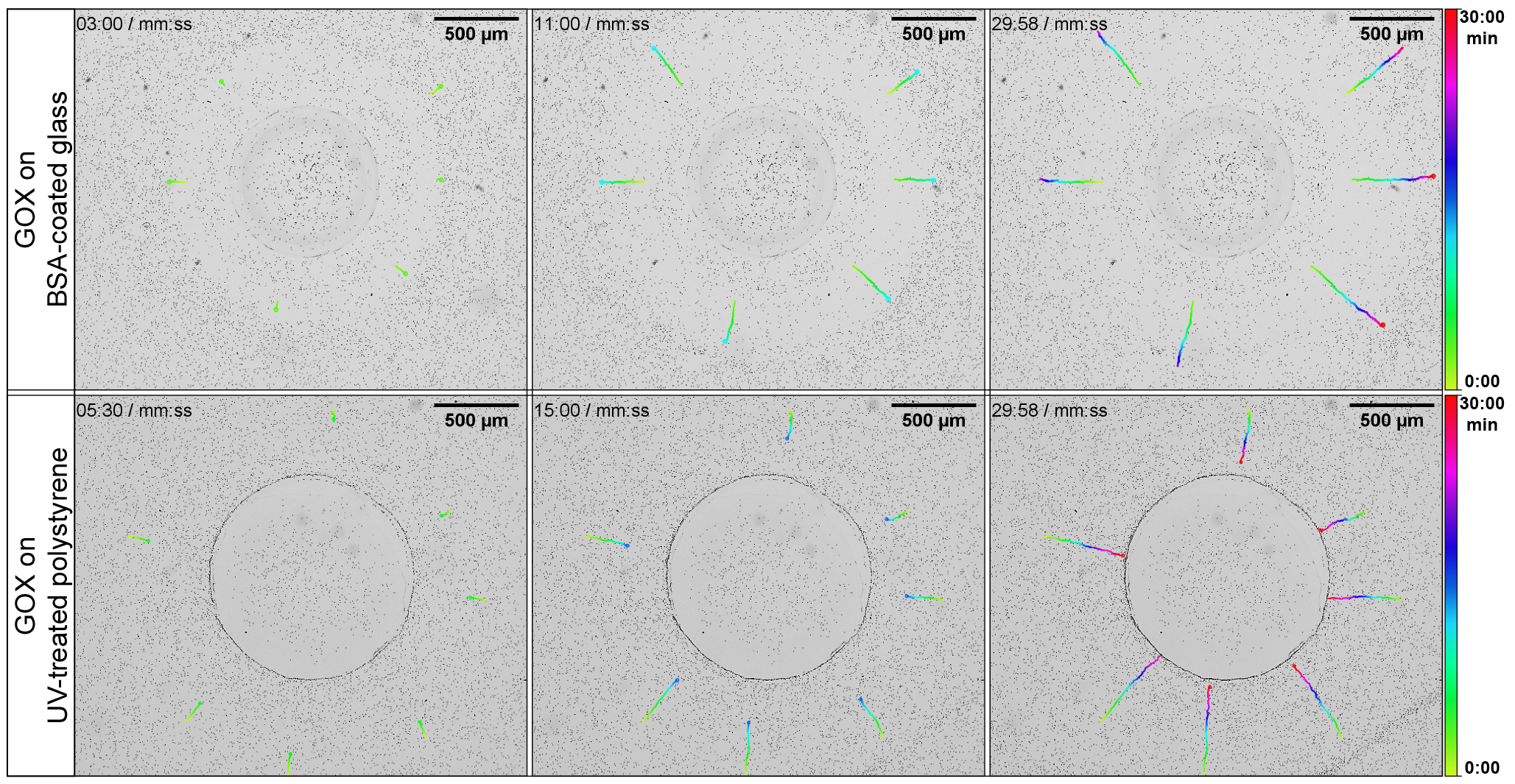}
\end{center}
\caption{
\label{fig:tracers_BSA_vs_PS}  Snapshots illustrating the time evolution of the distribution of tracers in the vicinity of GOX patches (the central disk-like shapes, delimited by slightly darker circular contours) imprinted onto either BSA-coated glass surface (upper row, see also the Supplementary Material Video S1) or UV-oxidized polystyrene surface (lower row, see also the Supplementary Material Video S2). The GOX patches are immersed in a solution containing 20 mM glucose, and thus are chemically active, and are surrounded by 3 $\mathrm{\mu m}$ diameter silica microspheres as tracer particles (there are also tracers that sedimented directly on top of the patch, as seen in the first column). The time along the tracked trajectories is color coded from yellow (short time) to red (long times) according to the legend bars at the right.
}
\end{figure*}
In the context of osmotic, surface driven flows, naturally emerges the question of what changes, if any, will occur when the planar wall, on which the active patch is imprinted, is changed. We have carried out exploratory experiments in this direction by using GOX enzyme patches, of a disk-like shape of diameter of approximately 1 mm, imprinted onto i.) bovine serum albumin (BSA)-coated glass surfaces (schematic \fref{fig:schematic_exper}(a)), ii.) UV-oxidized polystyrene surfaces (schematic  \fref{fig:schematic_exper}(b)), and iii.) UV-oxidized polystyrene surfaces which were subsequently also covered with BSA (schematic \fref{fig:schematic_exper}(c)). In all cases, the wall, on which the $\sim$ 1 mm diameter patch is imprinted, has a diameter of 1 cm, and the solution above the wall has a height of approximately 1~mm (see also \aref{sec:mater_meth}). The observable studied is the behavior of the distribution of tracer particles, sedimented in a monolayer above the wall, upon turning on the activity of the system by adding the glucose into the solution. As suitable tracers, heavy enough to rapidly sediment onto the planar wall carrying the GOX patch, we have used silica microspheres of 3 $\mathrm{\mu m}$ diameter; although there is a significant tendency of the tracers to adhere to the wall and to the patch (see the Supplementary Material videos), sufficiently many of them remain mobile to allow robust observations of tracer drift.

As can be inferred from the description above (see also  \fref{fig:schematic_exper}), the first two systems are characterized by the same patch but by a different support-wall, while the third system corresponds to a model system characterized by a same-nature, in terms of osmotic response, wall and patch (due to the BSA coating that covers both the catalytic patch and the wall). The GOX patches, regardless of the surface they are made onto and of being or not covered by BSA, catalyze the oxidation of glucose by oxygen and produce glucono-$\delta$-lactone and hydrogen peroxide. Then, the chemical activity leads to gradients in the densities of glucose, oxygen, glucono-$\delta$-lactone, and hydrogen peroxide, or at least in the last two (assuming that the glucose and oxygen are present in abundance). 

\begin{figure*}[!htb]
\begin{center}
\includegraphics[width=0.8\paperwidth]{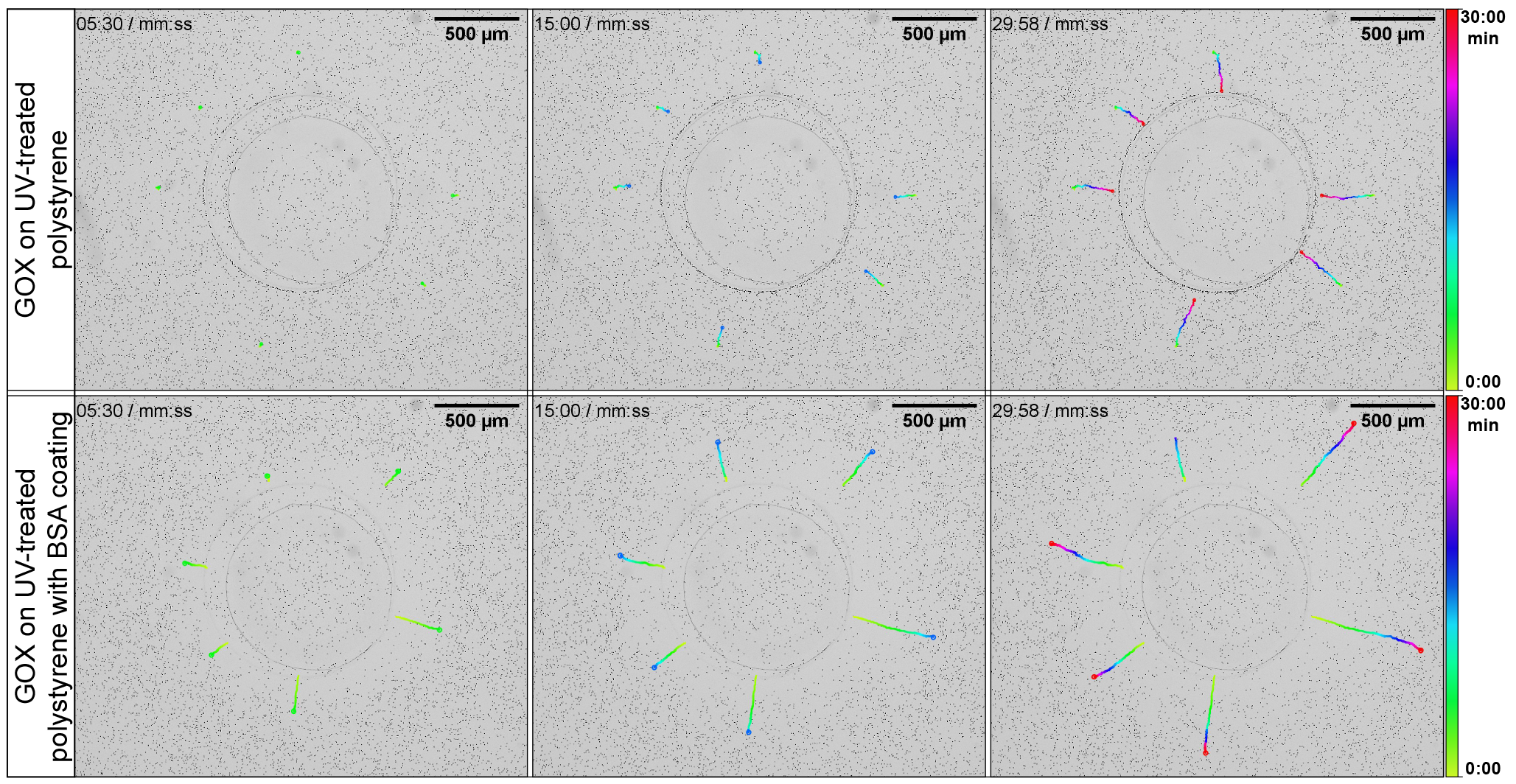}
\end{center}
\caption{
\label{fig:tracers_PS_vs_PSBSA}  Snapshots illustrating the time evolution of the distribution of tracers in the vicinity of GOX patches (the central disk-like shapes, delimited by slightly darker circular contours) imprinted onto an UV-oxidized polystyrene surface (upper row, see also the Supplementary Material Video S3) and then modified (after cleaning, see \aref{sec:mater_meth}) by application of a BSA coating over both the catalytic patch and the wall (lower row, see also the Supplementary Material Video S4). The GOX patches are immersed in a solution containing 20 mM glucose, and thus are chemically active, and are surrounded by 3 $\mathrm{\mu m}$ diameter silica microspheres as tracer particles (there are also tracers that sedimented directly on top of the patch, as seen in the first column). The time along the tracked trajectories is color coded from yellow (short time) to red (long times) according to the legend bars at the right. The double circle footprint of the patch, better visible in the top row, is most likely due to a slight lateral slipping of the ink-gel capillary tip during the printing of the patch.
}
\end{figure*}
\fref{fig:tracers_BSA_vs_PS} illustrate the evolution of the distribution of tracers within the monolayer for chemically active GOX patches (immersed in 
the glucose solution) imprinted onto a BSA-coated glass surface (top row) and onto a UV-oxidized polystyrene surface (bottom row) (the corresponding videos are provided in the Supplementary Material). While on the BSA-coated glass wall the monolayer particles around the patch exhibit an effective repulsion from the region of the patch, effectively creating a depletion zone that grows in time (see also Ref. \cite{munteanu_glucose_2019}), on the UV-oxidized polystyrene surface they exhibit an effective attraction towards the patch, leading to the dense accumulation at the rim of the patch.\footnote{As also indicated in the schematics in \fref{fig:schematic_exper}, the experimental patch is not flatly embedded in the wall, but has a small height (leading to the gray-contrast, visible in \fref{fig:tracers_BSA_vs_PS} and \fref{fig:tracers_PS_vs_PSBSA}, that allows identification of the region covered by the patch); this rim prevents the tracers from the outer region to move onto the patch.} (Although in the sequence of still snapshots the visibility of this accumulation is somewhat hindered  by a strong tendency of the tracers to stick to the polystyrene wall, the drift towards the rim is very clear in the video recordings of the tracers motion, see the Supplementary Material movies.) Since the composition of the solution is identical in the two cases, the chemical reaction is the same, and the tracer particles are the same, the only contribution --- among phoretic response, drift by bulk-driven flows, and drift by osmotic flows --- to the motion of the tracer that can change direction under these conditions is the one from the hydrodynamic flow due to induced osmotic slip at the wall. Therefore, the experiment provides a very clear demonstration that the osmotic flows have qualitative effects, discernible from other contributions, on the dynamics of the tracer particles that are close to the wall.

\fref{fig:tracers_PS_vs_PSBSA} illustrates the dynamics within the monolayer of tracers for chemically active GOX patches (immersed in the glucose solution) imprinted onto an UV-oxidized polystyrene surface (top row) and after coating both the patch and the wall with BSA (bottom row), see also the videos provided in the Supplementary Material. The dynamics in the monolayer in  the latter case is very similar with that in the case that only a glass wall, but not the patch, was covered by the BSA (top row in \fref{fig:tracers_BSA_vs_PS}). 
Taken together, these results suggests that, in these experiments, the osmotic flows are impacted by the surface-chemistry of the wall rather than by the surface-chemistry of the catalytic patch. However, one recalls that the  catalytic patches already contain BSA, and thus it is somewhat to be expected that the additional coating step impacts more the surface-chemistry of the wall than that of the patch.
 
\section{Model system}
\label{sec:model}

The discussion in the Introduction and in the previous section emphasized that both bulk-driven and osmotic, surface-driven effects can be present within an experimental realization of a chemically-active micropump. We proceed to an analytical calculation of the hydrodynamic flow driven by both type of sources by using a simple model system described below. While drastic simplifications --- from the complex experimental realizations --- such as: consideration solely of charge-neutral species, a simple ``constant flux'' reaction scheme, the use of a half space geometry, are made in order to allow analytical calculations, the resulting model allows us to obtain physical insight into the hydrodynamic flows induced by chemically active pumps as well as into the relative importance of each type of source of flow.   

\begin{figure}[!htb]
\includegraphics[width= \columnwidth]{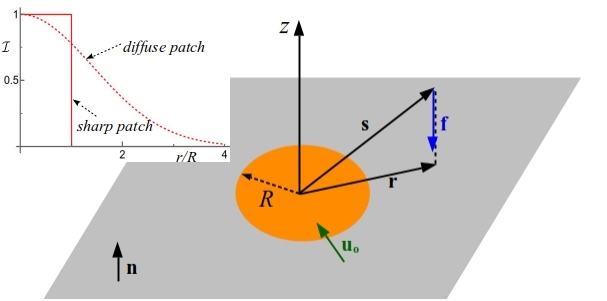
}
\caption{\label{fig_schematic} Schematic depiction (not to scale) of the 
geometry of the model system. A planar wall (the plane at $z = 0$) contains a chemically active patch (the orange region), here in the shape of a circular disk of radius $R$, of negligible thickness. The mixture (solvent, reactants, and products species), occupies the half-space $z > 0$. A generic point located within the solution at height $z$ and in plane position vector $\br$ is denoted as $\bs := (\br,z) = r \be_r + z \be_z$. The inner (into the fluid) normal unit vector at the wall, denoted by $\bn$, coincides with $\be_z$. The green arrow depicts the in-plane (at $z = 0^+$) osmotic slip vector field $\bu_\mathrm{o}(\br)$, while the blue arrow depicts the body-force density field $\bff(\bs)$ within the solution (see \sref{subsec:hydro}). The inset at the top left shows the two choices employed in this study for the radially-symmetric shape, described by $\mathcal{I}(r)$, of the chemically active patch. The step function shape (sharp patch) is typical for experimental studies, while the diffuse patch provides an example of a smooth, spatially unbounded (but sufficiently fast decaying with the distance) activity-distribution. 
}
\end{figure}
We model the system of interest as a dilute, ideal solution (consisting of solvent and solute molecules) occupying the half space $z > 0$ above a planar wall located at $z = 0$. The system is in contact with a reservoir which fixes the temperature and the bulk chemical potentials of the molecular components; under the assumption of an ideal solution, this implies that the average number density of solute far from the wall (in the bulk) is fixed to a constant value that we denote by $C_\infty$. On the wall there is an infinitesimally thin, chemically active patch (see \fref{fig_schematic}). The chemical activity at the patch is modeled as a source (or sink) of solute, i.e., at the patch solute molecules are released (removed) from the solution. The solute is assumed to diffuse freely in the solution with diffusion constant $D$ and to behave, to a good approximation, as an ideal gas. Motivated by the typical experimental realization of the active patch being that of a catalyst promoting a chemical reaction within the solution, we further assume that (i) the reaction providing the chemical activity is not affected by the transport of reactant molecules --- which here are implicitly considered as part of the solvent --- to the patch (i.e., at small Damk{\"o}hler number, as the catalytic conversion kinetics is assumed to be reaction-rate limited), and (ii) that the solute is efficiently transported away from the patch towards the bulk solution, i.e., we disregard any possibility of a catalyst poisoning. Accordingly, the activity of the patch is characterized by a time-independent rate $Q$ (strength of the source/sink) of release/annihilation of solute molecules per unit area of the patch; for simplicity, $Q$ is further assumed to also be spatially constant over the area of the patch. 

An equilibrium distribution of solute in solution is dictated by the reservoir (which maintains the homogeneous $C_\infty$ far from the wall) and the adsorption potential of the wall (manifested as an excess or a depletion of solute in the vicinity of the wall due to different interactions wall-solute molecule and wall-solvent molecule); the presence of the active patch drives the system out of equilibrium by perturbing the solute distribution. In the following, we assume that a steady state of the fluid solution, characterized by a distribution $c(\bs)$ of the number density of the solute and a hydrodynamic flow $\bu(\bs)$ of the solution (where $\bs$ denotes a point within the solution, see Fig. \ref{fig_schematic}), emerges, and we aim at determining these two fields. We restrict the analysis to the case that the patch has a disk shape, and thus the whole system has axial symmetry; accordingly, we look for axisymmetric solutions for the two fields (which, in terms of the cylindrical coordinates, will depend on $r$ and $z$, but not on the azimuthal angle).

\section{Mathematical formulation}
\label{sec:calc_steady_state}

\subsection{Solute distribution}
\label{subsec:solute_dist}

Since the solute is conserved in the bulk solution (i.e., there are no sources or sinks at $z > 0$), the distribution of the solute at steady state obeys $\nabla \cdot \bj = 0$ in terms of the current $\bj$ of solute molecules. Under the assumption that the diffusion of the solute is fast while the hydrodynamic flow is slow, such that the transport of the species by diffusion dominates over the one by advection (i.e., the transport P{\'e}clet number is very small), the current of solvent is diffusive,\footnote{Actually, this holds only in the region beyond the range $\Lambda$ of the adsorption potential; by assuming that $\Lambda \ll R$, which is the case for the typical experimental realization, and anticipating that the lengthscale of the variations of $c$ is of the order of $R$, one can replace the analysis within the very thin boundary layer region near the wall with a boundary condition, \eq{BC_n_wall} below, at the wall \cite{MiLa14}.} $\bj = - D \nabla c$.
Consequently, one infers that the distribution of solute at steady state obeys the Laplace equation 
\begin{equation}
 \label{Lapl_n}
\nabla^2 c(\bs) = 0
\,.
\end{equation}  
The solute distribution is subject to the boundary conditions (BC) at infinity (bulk solution),
\begin{subequations}
 \label{BCs_n}
 \begin{equation}
 \label{BC_n_infty} 
 c(|\mathbf{s}| \to \infty) = C_\infty\,,
\end{equation}
and on the wall, 
 \begin{equation}
 \label{BC_n_wall} 
\left.\lbrace\mathbf{n} \cdot [-D \nabla c(\mathbf{s})]\rbrace\right|_{z = 0} = 
 Q \,{\cal I}(r)\,,
\end{equation}
\end{subequations}
respectively. In Eq. (\ref{BC_n_wall}), $\mathbf{n} = \mathbf{e}_z$ denotes the inner normal (into the fluid) at the wall (see Fig. \ref{fig_schematic}), while the dimensionless $\mathcal{I}(r) \geq 0$ denotes the activity-distribution function, which describes the geometry (position, shape, extent) of the patch as well as the local magnitude of the activity. Eq. (\ref{BC_n_wall}) defines the chemical activity of the patch (release/removal of solute) as a diffusion current, along the normal direction, into the solution, the value of which is prescribed to be equal to the rate $Q {\cal I}(r)$ of the chemical activity of the patch (accordingly, $Q > 0$ corresponds to a source, while $Q < 0$ to a sink). Regarding  $\mathcal{I}(r) \geq 0$, we require that it decays sufficiently fast with $r$ (at least faster than $r^{-2}$, such that the monopole at the chemically active wall is finite), and we fix its maximum to 1 (which is always possible by a suitable choice for $Q$). Further restricting the discussion to the case that $\mathcal{I}(r)$ is monotonic, we set $\mathcal{I}(0) = 1$.  

\subsection{Hydrodynamic flow}
\label{subsec:hydro}

The hydrodynamic flow $\bu(\bs)$ is studied under the assumption that the solution behaves as a Newtonian fluid. The viscosity $\mu$ is assumed to be spatially constant, even though the chemical composition  of the solution is spatially varying (recall that $c(\bs)$ is a function of position). The flow is assumed to be incompressible and (motivated by the typical flow velocities reported in experimental studies) to be of very small Reynolds number; it thus obeys the creeping flow Stokes equations. The hydrodynamics is driven by the chemical activity at the patch on the wall, which manifests as two distinct sources for the flow. These sources are discussed below, and the boundary value problem obeyed by the hydrodynamic flow is summarized in the last part of this Subsection.

\subsubsection{Actuation at the wall as an active osmotic slip}
\label{subsubsec:sli_wall}
The first source of hydrodynamic flow is an ``osmotic slip'' actuation at the wall \cite{DSZK47,Ande89},
\begin{equation}
 \label{eq:slip_vel}
 \bu_o(\br) = -b(\br) \nabla_\parallel c(\br)\,, 
\end{equation}
which is due to the spatial variations (generated by the chemical activity at the patch) in the number density of solute along the wall and to a spatially extended --- but of microscopic range $\Lambda \ll R$ --- ``adsorption potential'' exerted by the wall on the solute molecules. In \eq{eq:slip_vel}, $\nabla_\parallel$ denotes the gradient operator projected along the surface of the wall, while 
$b(\br)$, which is determined by the adsorption potential \cite{DSZK47,Ande89,MiLa14}, denotes the so-called osmotic slip coefficient; the minus sign is introduced for historical reasons, following the original sign convention in Refs. \cite{DSZK47,Ande89}.  Rigorously speaking, the tangential derivative in \eq{eq:slip_vel} is to be evaluated at some $z = \sigma$, with 
$\Lambda \lesssim \sigma \ll R$, outside of the thin layer defined by the range of the adsorption potential; i.e., for the flow, which is expected to vary over length scales of the order of $R$, at $z = 0^+$ rather than directly at the boundary $z = 0$ \cite{Ande89}. (While in many cases this difference is in practice insignificant, one of the instances of relevance is the one, discussed latter in the text, of a chemically active patch with the sharp-edge shape given by the Heaviside step function.)

If the material composition of the surface is heterogeneous (which is the typical case for a patch of catalytic material imprinted on a chemically inactive wall), the coefficient $b(\br)$ in general varies along the surface; for simplicity, here the spatial-dependence of $b(\br)$ is taken to be the same as that of the chemical activity. Denoting by $b_w$ the value at the wall (i.e., far from the patch, where $\mathcal {I} \to 0$) and by $b_c$ the value at $r = 0$ (where the chemical activity is maximal) we write\footnote{At least one of the values $b_c$ and $b_w$ is non-zero -- otherwise there is no surface-driven flow due to the active patch active actuation at the wall. Without lack of generality, we assumed $b_w \neq 0$ when defining the parameters $\beta$; if that is not the case, then necessarily $b_c \neq 0$ and one simply redefines the $\beta_{c,w}$ in units of $|b_c|$ instead of $|b_w|$.}
\begin{eqnarray}
 \label{eq:def_b}
b(r)&=& b_w + (b_c - b_w) \mathcal {I}(r) \\
&:=& |b_w| [\beta_w + \underbrace{(\beta_c - \beta_w)}_{\hspace*{.5cm}:= \,\delta \beta} \mathcal {I}(r)]:= |b_w| \beta(r)   \nonumber
\end{eqnarray}
in terms of $\beta_{c,w} := b_{c,w}/|b_w|$.

\subsubsection{Body-force distribution within the solution}
\label{subsubsec:body_forces}
The second source for the hydrodynamic flow induced by the out of equilibrium, spatially non-uniform distribution of solute stems out from the coupling of the variations in the mass density of the solution with gravity (the so-called "solutal buoyancy" \cite{valdez_solutal_2017}). By assuming that these variations in mass density are small in comparison to the average mass density of the bulk solution (e.g., Ref. \cite{valdez_solutal_2017} estimates them to be less than 0.3\% of the average density), their effect can be modeled as a distribution of body forces $\bff(\bs)$ acting on a fluid of uniform (average) density (the Boussinesq approximation, see, e.g., Ref. \cite{Barletta_2022}), which implies incompressibility of the flow. Since the variations $\Delta \rho$ in the mass density are due to the variations in the number density of the solute around the bulk (equilibrium) value $C_\infty$, one can write $\Delta \rho (\bs) = \alpha\, 
[c(\bs) - C_\infty]$ with some proportionality constant $\alpha$ accounting also for the correct dimensions. This renders for the distribution of body forces the expression
\begin{equation}
\label{eq:f_dist}
\bff(\bs) = \alpha\, [c(\bs) - C_\infty]\, \bg\,,
\end{equation}
where $\bg$ denotes the gravitational acceleration. In the following, we will consider only the case in which the system is aligned such that the wall is horizontal, i.e., $\bg = g \be_z$, and that $g < 0$, corresponding to the fluid being situated \textit{above} the wall on which the patch is imprinted (note that the case $g > 0$ is equivalent to a system in which the fluid is situated \textit{below} the wall containing the patch).

\subsubsection{The hydrodynamics boundary value problem}
\label{subsubsec:Stokes_bvp}
Putting everything together, the hydrodynamic flow $\bu(\bs)$ is obtained as the solution of the incompressible Stokes equations
\begin{equation}
\label{eq:Stokes}
-\nabla p + \mu \nabla^2 \bu + \bff = 0\,,~~~ \nabla \cdot \bu = 0\,,
\end{equation}
where $p(\bs)$ denotes the pressure field which ensures the incompressibility condition and $\bff(\bs)$ is the body force density, \eq{eq:f_dist}, 
subject to the boundary conditions of quiescent flow at infinity (far from the patch),
\begin{subequations}
 \label{eq:BCs_flow}
 \begin{equation}
 \label{eq:BC_u_infty} 
 \bu(|\mathbf{s}| \to \infty) = 0\,,
\end{equation}
and of osmotic slip, \eq{eq:slip_vel}, at the wall 
 \begin{equation}
 \label{eq:BC_flow_wall} 
\bu(\br,z = 0) = \bu_o(\br)\,.
\end{equation}
Due to the linearity of the equations and of the BCs, the solution will be constructed by superposition of two flows, $\bu = \bv + \bw$. The flow $\bv$ is the solution of the homogeneous, incompressible Stokes equations ($\bff$ set to zero in \eq{eq:Stokes}) subject to osmotic slip and quiescent flow BCs as in \eqs{eq:BCs_flow}. Then, accordingly, the flow $\bw$ is the solution of the inhomogeneous \eq{eq:Stokes} obeying \eq{eq:BC_u_infty} as BC at infinity and with a no-slip BC at the wall, i.e.,   
 \begin{equation}
 \label{eq:BC_w_wall} 
\bw(\br,z = 0) = 0\,,
\end{equation}
\end{subequations}
replacing \eq{eq:BC_flow_wall}. The flows $\bv$ and $\bw$ will be calculated in the following sections, after discussing the solution to the diffusion equation for the solute.

\section{The steady state distribution of solute}
\label{sec:distri_solute_gen_sol}
The general axisymmetric solution of \eq{Lapl_n} satisfying the BC at infinity, \eq{BC_n_infty}, can be written as \cite{Kim_1983}
\begin{equation}
\label{eq:C_gen}
c(\bar{r}, \bar{z}) = C_\infty + C_0 
\underbrace{\int\limits_0^\infty d\xi \,\mathfrak{c}(\xi)\, \mathrm{J}_0\left(\xi \bar{r} \right)\,e^{-\xi \bar{z}}}_{:= { \bar c} (\bar{r},\bar{z})}\,,
\end{equation}
where $\bar{r} = r/R$, $\bar{z} = z/R$, and ${ \bar c} (\bar{r},\bar{z})$ are dimensionless. The function $\mathfrak{c}(\xi)$ is determined from the BC at the wall, \eq{BC_n_wall}, which renders
\begin{equation}
\label{eq:BC_wall_for coef_c_gen}
\int\limits_0^\infty d\xi \,\xi \,\mathfrak{c}(\xi)\, \mathrm{J}_0\left(\xi \bar{r} \right) = 
\frac{Q R}{D C_0} \mathcal{I}(\bar r)\,.
\end{equation}
By choosing the density scale $C_0$ as
\begin{equation}
 \label{eq:dens_scale}
 C_0 : = \frac{Q R}{D}
\end{equation}
and employing the orthogonality of the Bessel functions of given index $n$ \cite{Abram_book}
\begin{equation}
\label{eq:ort_Bessel}
\int\limits_0^\infty d\chi \,\chi \,\mathrm{J}_n\left(\xi \chi\right)\mathrm{J}_n\left(\xi' \chi \right)= \xi^{-1} \delta\left(\xi- \xi'\right)\,,
\end{equation}
\eq{eq:BC_wall_for coef_c_gen} leads to 
\begin{equation}
\label{eq:coef_c_gen}
\mathfrak{c}(\xi) = \int\limits_0^\infty \, d\bar{r} \,\bar{r} \,\mathcal{I}(\bar{r}) \, \mathrm{J}_0\left(\xi \bar{r} \right)\,.
\end{equation}
This concludes the calculation of the solution of the diffusion problem in  \eqs{Lapl_n} and \eref{BCs_n} for a generic axisymmetric activity function $\mathcal{I}(r)$.

\section{The flow due to the active osmotic slip}
\label{sec:osmo_hydro}

By using the general form of the axisymmetric flow solution of the homogeneous, incompressible Stokes equation provided by Ref. \cite{Kim_1983} (see also \cite{Menzel_2020}), and using the BC 
at infinity to exclude terms proportional to $\exp(\xi \bar{z})$, the flow $\bv = v_r \be_r + v_z \be_z$ is written as
\begin{subequations}
\label{eq:v_start}
\begin{equation}
\label{eq:vr_start}
\frac{v_r(\bar{r}, \bar{z})}{V_0}= \int\limits_0^\infty d\xi \,\xi \left[A(\xi) \bar{z} + B(\xi)\right] \,e^{-\xi \bar{z}}\, \mathrm{J}_1\left(\xi \bar{r} \right)\,,
\end{equation}
\begin{equation}
\label{eq:vz_start}
\frac{v_z(\bar{r}, \bar{z})}{V_0} = \int\limits_0^\infty d\xi \left[A(\xi) \left(1+ \xi \bar{z}\right) + \xi B(\xi) \right] \,e^{-\xi \bar{z}}\,\mathrm{J}_0\left(\xi \bar{r}\right)\,,
\end{equation}
\end{subequations} 
where the characteristic velocity is chosen as
\begin{equation}
 \label{eq:vel_scale}
 V_0 : = \frac{Q |b_w|}{D}\,,
\end{equation}
while the functions $A(\xi)$ and $B(\xi)$ are to be determined from the BC at the wall (at the edge of the thin layer). The later involves the osmotic slip, \eq{eq:slip_vel}, which, by using \eq{eq:C_gen} for the solute density, is given by
\begin{equation}
 \label{eq:slip_vel_V0}
 \bu_o(\bar r) 
 = V_0 \beta(\bar r)
\underbrace{ 
 \left(\int\limits_{0}^{\infty} \,
 d \xi \, \xi \,\mathfrak{c}(\xi) \,\mathrm{J}_1 (\xi \bar r) \,e^{-\xi \,\sigma} 
 \right) 
 }_{:= {\bar u}_o({\bar r};\sigma)}
 \be_r \,, 
\end{equation}
where the parameter $\sigma$ (a distance in units of $R$) is a reminder that the tangential derivative in the osmotic slip may have to be evaluated at a point slightly above the wall ($z = 0^+$, thus $0 < \sigma \ll 1$).

The projection along $\be_z$ of the BC at the wall, \eq{eq:BC_flow_wall}, 
implies $v_z(r,z=0) = 0$; accordingly, \eq{eq:vz_start} combined with the orthogonality relation \eq{eq:ort_Bessel}, renders 
\begin{equation}
\label{eq:A_func}
A(\xi) =  - \xi B(\xi)\,. 
\end{equation}
The projection along $\be_r$ of the BC at the wall, \eq{eq:BC_flow_wall},
combined with $v_r(r,z=0)$ from \eq{eq:v_start}, the orthogonality relation \eq{eq:ort_Bessel}, the expression of the slip velocity in 
\eq{eq:slip_vel_V0}, and the expression \eq{eq:def_b} of the osmotic slip coefficient, renders
\begin{eqnarray}
\label{eq:B_func}
B(\xi) &=& \int\limits_0^\infty d{\omega}\, {\omega}\, \beta({\omega}) \,{\bar u}_o({\omega};\sigma) \,\mathrm{J}_1\left(\xi \omega\right) \nonumber\\
&:=& \beta_w \, \mathfrak{b}_1(\xi;\sigma) + \delta \beta \,\mathfrak{b}_2(\xi;\sigma) \,.
\end{eqnarray}  
By using \eq{eq:slip_vel_V0}, the functions $\mathfrak{b}_{1,2}(\xi;\sigma)$ are expressed in terms of the coefficients $\mathfrak{c}(\xi)$, \eq{eq:coef_c_gen}, as 
\begin{subequations}
\label{eq:b_func}
\begin{equation}
\label{eq:b1_func}
\mathfrak{b}_1(\xi;\sigma) = \mathfrak{c}(\xi) \,e^{-\xi \,\sigma}\,,
\end{equation}
when using the orthogonality relation in \eq{eq:ort_Bessel}, and
\begin{equation}
\label{eq:b2_func}
\mathfrak{b}_2(\xi;\sigma) = \int\limits_0^\infty d{\omega}\, {\omega}\, \mathcal{I}(\omega) \,\mathrm{J}_1\left(\xi \omega\right) q(\omega;\sigma)\,,
\end{equation}
where 
\begin{equation}
q(\omega;\sigma) = \int\limits_0^\infty d\chi \,\chi \,\mathfrak{c}(\chi)\, \mathrm{J}_1\left(\chi \omega\right) \,e^{-\chi \,\sigma}\,.
\end{equation}
\end{subequations}
Collecting the results above, one concludes that the flow $\bv(\bs)$ is given by the superposition
\begin{equation}
\label{eq:v_v1_v2}
\bv(\bs) = V_0 [\beta_w {\bv}_1(\bar \bs) + \delta \beta {\bv}_2(\bar \bs)]
\end{equation}
of the base flows 
\begin{eqnarray}
\label{eq:v1_v2}
{\bv}_{1,2}(\bar\bs) &:=& \be_r \int\limits_0^\infty d\xi \,\xi 
(1-\xi \bar z) \mathfrak{b}_{1,2}(\xi;\sigma)
\,e^{-\xi \bar{z}}\, \mathrm{J}_1\left(\xi \bar{r} \right) \nonumber \\
&-& \be_z \,\bar z \int\limits_0^\infty d\xi \,\xi^2 
\mathfrak{b}_{1,2}(\xi;\sigma)
\,e^{-\xi \bar{z}}\, \mathrm{J}_0\left(\xi \bar{r} \right)\,,
\end{eqnarray}
where $\bar \bs:= \bs/R$. \eqs{eq:slip_vel_V0} and \eref{eq:def_b} provide a
straightforward interpretation of the superposition above: ${\bv}_{1}$ is the hydrodynamic flow that would occur for a wall that has a uniform osmotic coefficient, while ${\bv}_{2}$ is the flow induced by an osmotic slip distribution that is non-zero only in the region of the patch. This concludes the calculation of the flow $\bv(\bs)$ for a generic axisymmetric activity function $\mathcal{I}(r)$ and an osmotic slip coefficient as in \eq{eq:def_b}. 

\section{Example: the case of a compact support active patch.}
\label{sec:step_funct_activ}

\begin{figure}[!htb]
\includegraphics[width = \columnwidth]{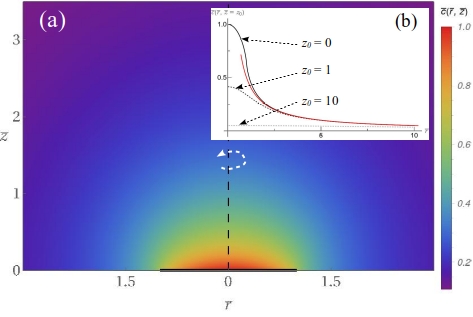}
\caption{\label{fig:sol_dens_compact} (a) Color coded (bar scale at the right) dimensionless density of solute ${\bar c}(\bar r, \bar z)$, \eqs{eq:C_gen} and \eref{eq:c_compact_patch}, for the case of a compact, sharp edged patch (\eq{eq:compact_patch}); the thick black segment at $\bar z = 0$ depicts the location and extent of the active patch, while the curved white arrow is a reminder of the axial symmetry around O$z$. (b) In plane ($\bar z = z_0$) dimensionless density of solute ${\bar c}(\bar r,z_0)$ at various heights $\bar z = z_0$ above the wall. The red line shows the far field (monopolar) behavior $\sim {\bar r}^{-1}$ for ${\bar c}(\bar r,z_0)$ at fixed $z_0$.
}
\end{figure}

A typical experimental realization is that of a deposition or imprinting of the catalyst material 
within a well defined disk like region of the wall. This is modeled here via an activity function given by the Heaviside step function,
\begin{equation}
 \label{eq:compact_patch}
\mathcal{I}(\bar r) = 
\begin{cases} 
1, & \bar r \leq 1\, \\
0,  & \bar r > 1\, 
\end{cases}\,.
\end{equation}
For this activity function, \eq{eq:coef_c_gen} renders (see also Ref. \cite[Ch.~3.2.3-4]{Sneddon_book})
\begin{equation}
\label{eq:c_compact_patch}
\mathfrak{c}(\xi) = \xi^{-1} \mathrm{J}_1(\xi)
\end{equation}
for the coefficients in the integral representation of the density. The integral representation cannot be further simplified; accordingly the density $\bar c(\bar r,\bar z)$ is calculated by performing the integral numerically, using the software Mathematica (version 13) at any given point $\bar \bs = (\bar r,\bar z)$. The result, shown in \fref{fig:sol_dens_compact}, exhibits the expected $1/|\bs|$ asymptotic decay with the distance from the patch (corresponding to the far field interpretation as the diffusion field from a monopolar source at the origin).

\begin{figure}[!htb]
\includegraphics[width=0.95\columnwidth]{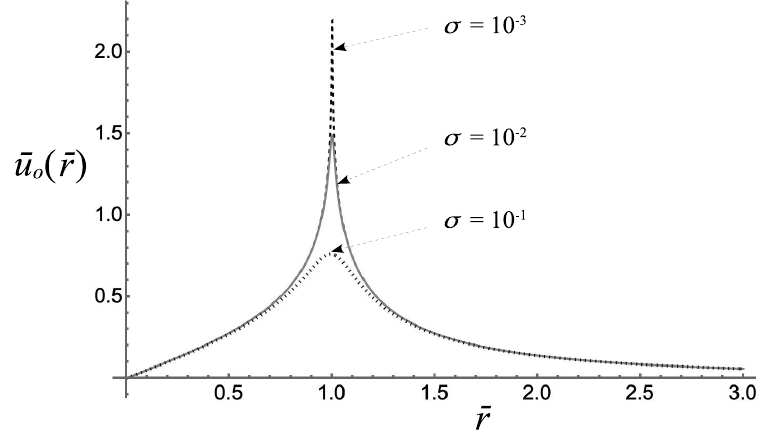}
\caption{\label{fig:slip_sig} The dimensionless osmotic slip ${\bar u}_0(\bar r)$, \eq{eq:slip_vel_V0},  for several values of the parameter $\sigma$. 
}
\end{figure}

As anticipated in \sref{subsubsec:sli_wall}, for this choice of activity function the osmotic slip cannot be evaluated at $z=0$: for $\sigma =0$, \eq{eq:slip_vel_V0} with the coefficient $\mathfrak{c}(\xi)$ above has a cusp divergence at $r = R$, i.e., at the \textit{sharp} edge of the patch. This isolated singularity, which here we remove by evaluating the osmotic slip at $\bar z = \sigma$ with $0 < \sigma \ll 1$, see \fref{fig:slip_sig}, is an artifact of the sharp-edge assumption. 
It can alternatively be removed, with the same effect on the osmotic slip as in \fref{fig:slip_sig}, by considering that the lateral border of the patch has a small, but non-vanishing, width $\epsilon$ (in units of $R$) within which the fast, but smooth, decay of the activity happens. It turns out that: (i) the flows ${\bv}_{1,2}(\bar\bs)$ depend only quantitatively, but not qualitatively, on the value of $\sigma$; (ii) the use of a model with an interface of non-zero width also introduces an additional model parameter, $\epsilon$, which has similar influences on the results (i.e., regularizes the cusp divergence and influences the magnitude of the base flows ${\bv}_{1,2}(\bar\bs)$, but not their qualitative features) while requiring more involved algebra; and (iii) even for a diffuse patch (i.e., $\mathcal{I}(r)$ does not have a compact support, but decays sufficiently fast with the distance from the patch, as shown in \fref{fig_schematic}), where there are no mathematical issues with the behavior exactly at the wall, the results are qualitatively similar (see \aref{sec:diffuse_patch}). Because of these, we will 
proceed here with the model of a compact patch with sharp edge, \eq{eq:compact_patch}, and illustrate the resulting flows for $\sigma = 10^{-3}$.

\begin{figure}[!t]
\includegraphics[width=0.9\columnwidth]{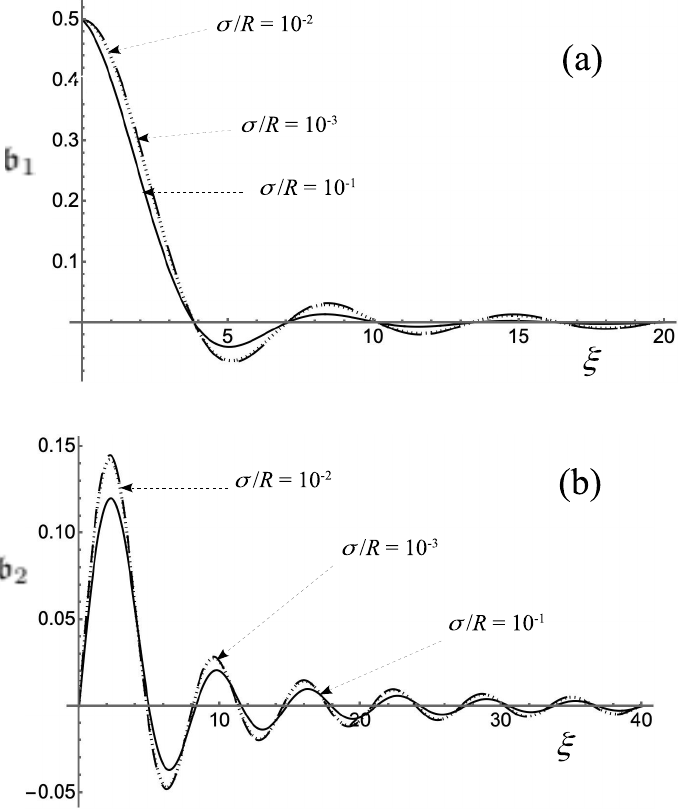}\hfill\\
\hspace*{1.5cm}$\sigma/R =$~~~$10^{-1}$ (---) ~~~ 
$10^{-2}$~($\cdots$)~~~$10^{-3}$ (- - -) \hfill
\caption{\label{fig:b1_and_b2_compact} The coefficients $\mathfrak{b}_1(\xi)$ (a) and $\mathfrak{b}_2(\xi)$ (b) as functions of $\xi$ for several values of the parameter $\sigma$. The results at $\sigma = 10^{-2}$ and $\sigma = 10^{-3}$ are practically indistinguishable.
}
\end{figure}

\begin{figure}[!tb]
\includegraphics[width=0.99\columnwidth]{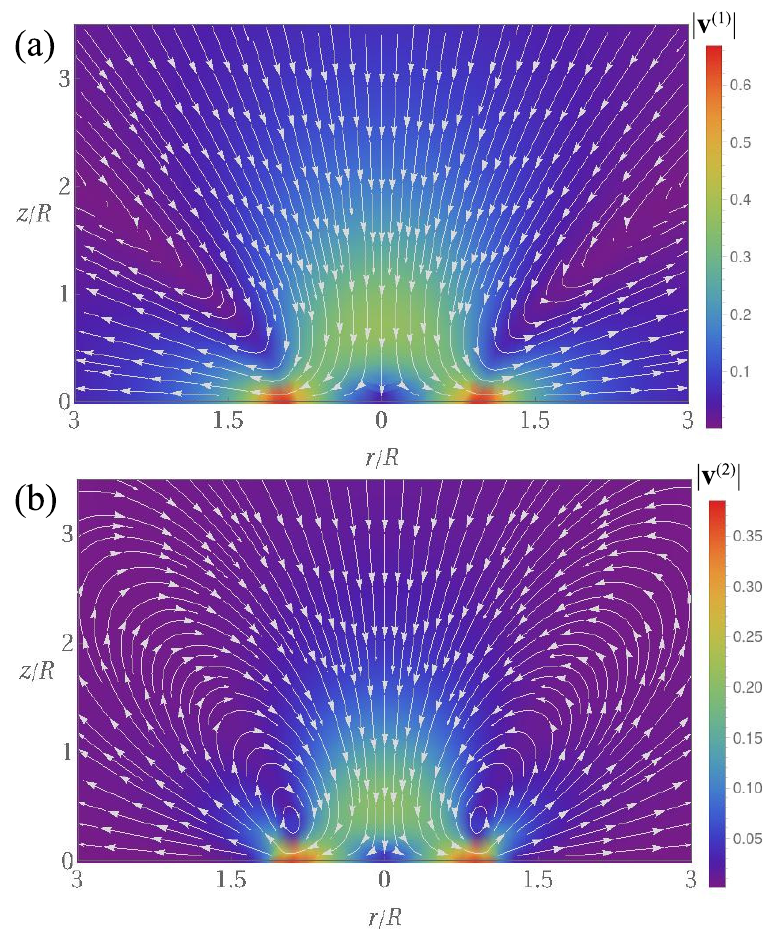}
\caption{\label{fig:v1_and_v2_compact} Streamlines (white) and magnitude (color coded background) for the basic flows $\bv^{(1)}(\bar \bs)$ (panel (a)) and $\bv^{(2)}(\bar \bs)$ (panel (b)) driven by the osmotic slip induced by the active patch (\eqs{eq:v1_v2}, \eref{eq:b_func}, and \eref{eq:c_compact_patch}). The parameter $\sigma$ is fixed to the value $\sigma = 10^{-3}$.   
}
\end{figure}

By \eqs{eq:b1_func} and \eref{eq:c_compact_patch}, the function $\mathfrak{b}_1(\xi;\sigma)$ (shown in \fref{fig:b1_and_b2_compact}(a) for several values of $\sigma$) is known in explicit form (which allows one to infer that the main influence of the parameter $\sigma$ is to induce an exponential upper cut-off for the $\xi$-dependence); however, the integral representation of the flow ${\bv}_{1}(\bar\bs)$ cannot be further simplified. 
Accordingly, as in the case of the solute density, the flow ${\bv}_{1}(\bar\bs)$ at a given point $\bar \bs$ is calculated by evaluating numerically the corresponding integrals in \eq{eq:v1_v2} using the software Mathematica 13. The result of such a computation on a rectangular grid in the range $0\leq \bar r \leq 3$ and $5 \times 10^{-2} \leq \bar z \leq 3.5$ is shown in  \fref{fig:v1_and_v2_compact}(a) as white streamlines on a color background that encodes the magnitude $|{\bv}_{1}|$ of the flow.  

The calculation of the flow $\bv^{(2)}(\bs)$ is more involved in that the 
integral over $\omega$ in \eq{eq:b2_func} cannot be calculated analytically, and thus the coefficient $\mathfrak{b}_2(\xi;\sigma)$ remains given as an integral representation\footnote{One notes that $\mathfrak{b}_2(\xi;0)$ can be computed exactly in terms of the hypergeometric function $_2F_3$ \citep{Abram_book}. One also notes that the functions $\mathfrak{b}_{1,2}(\xi;\sigma)$ are finite for $\sigma=0$, and the corresponding integral representations of the velocity fields $\bv_{1,2}$ are convergent, including at $z=0$ in the sense of distributions, when $\sigma \to 0$.}
\begin{equation}
\label{eq:b2_compact_patch}
\mathfrak{b}_2(\xi;\sigma) = \int\limits_0^\infty d\omega e^{-\omega \,\sigma} 
\mathrm{J}_1(\omega) 
\frac{\xi \mathrm{J}_0(\xi) \mathrm{J}_1(\omega) - \omega \mathrm{J}_0(\omega)\mathrm{J}_1(\xi)}{\omega^2-\xi^2}\,.
\end{equation}
Accordingly, in this case we set an upper cut-off $\xi_M = 100$ (which is sufficient to ensure convergence of their numerical computation at the smallest $\bar z$, $\bar z = 5 \times 10^{-2}$, of the grid of coordinates considered here) and calculate the coefficients $\mathfrak{b}_2(\xi)$ at a list of $10^3$ values $\xi_k$ uniformly spaced in the interval $0 \leq \xi \leq \xi_M$; then $\mathfrak{b}_2(\xi)$ at any  $0\leq \xi \leq \xi_M$ is obtained by interpolation of the pre-calculated values $\mathfrak{b}_2(\xi_k)$. The results $\mathfrak{b}_2(\xi)$ thus obtained are shown in \fref{fig:b1_and_b2_compact}(b) for several values of $\sigma$. The resulting interpolating function $\mathfrak{b}_2(\xi)$ is then plugged in the numerical computation (with Mathematica 13) of the integrals involved in the flow velocities, the range of integration being also restricted to $0\leq \xi \leq \xi_M$. The resulting flow $\bv^{(2)}(\bar \bs)$ is shown in \fref{fig:v1_and_v2_compact}(b) (white streamlines, the color background encodes the magnitude $|{\bv}_{2}|$ of the flow). It can be seen that its magnitude is of the same order as that of $\bv^{(1)}(\bs)$; but the most interesting aspect is the occurrence of a closed vortex, extending significantly into the fluid, near the edge of the patch. This is interesting because the location and the extent of the vortex place it within the region typically explored by the experimental observations of the tracer motion. Such feature would be very difficult to discriminate experimentally from a ``convection roll'', as expected to occur in the case of bulk-driven flow within a closed vessel (see, e.g., Refs. \cite{valdez_solutal_2017,ortiz-rivera_convective_2016}), which further emphasizes the need for a careful interpretation of the experimental observations of tracer motion and, when possible, additional cross-validating experiments.

\begin{figure*}[!htb]
\begin{center}
\includegraphics[width=0.99\textwidth]{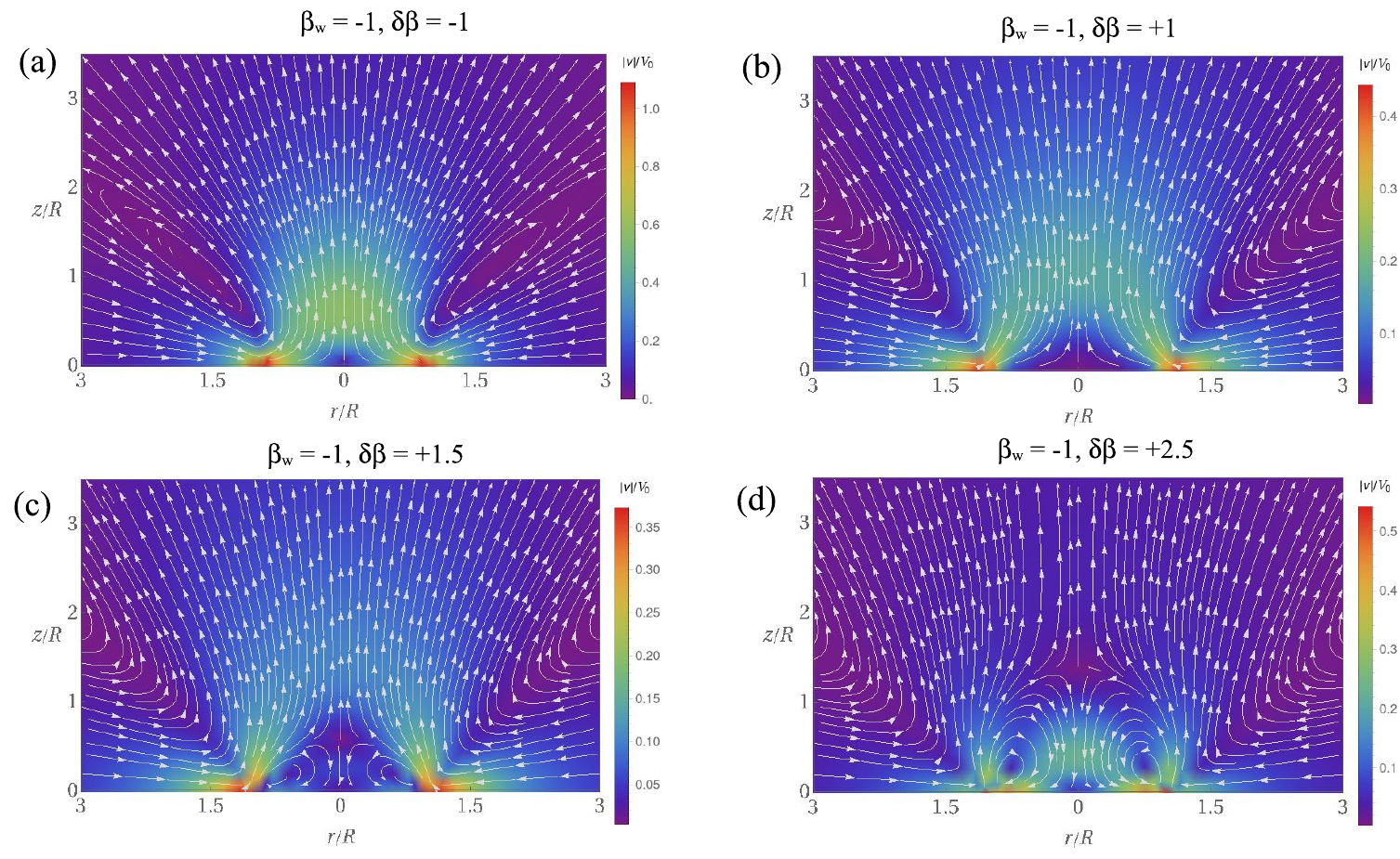}
\end{center}
\caption{
\label{fig:flow_v_b_and_db} Streamlines (white) and magnitude (color coded background) of the flow $\bv(\bs)/V_0 = \beta_w \bv^{(1)}(\bs) + \delta \beta \bv^{(2)}(\bs)$ (\eqs{eq:v1_v2}, \eref{eq:b_func}, and \eref{eq:c_compact_patch} with $\sigma =  10^{-3}$, for $\beta_w = -1$ and several values of $\delta \beta$. 
}
\end{figure*}
Finally, it is instructive and insightful to consider the full flow $\bv$ as a function of the amplitudes $\beta_w$ and $\delta \beta$ in the superposition of the two basic flows. This could be experimentally realized, e.g., by imprinting a catalyst patch over slides of different material (glass, polystyrene, etc), as discussed by us in \sref{sec:exper_motiv} (see Ref. \cite{farniya_imaging_2013} for a similar set-up involving a more complex chemically active bimetal redox system), or by using patches of different types of catalysts imprinted on the same type of wall, e.g., glass slides; the first alternative is preferable because the latter has the disadvantage that, in general, one would change at the same time the type of solution and the type of chemical reaction involved in the activity.

Since $\beta_w$ can take only the values $\pm 1$, we select one of them, $\beta_w = -1$ (``repulsive'' interaction between the wall and the solute), and vary the parameter $\delta \beta$. As shown in \fref{fig:flow_v_b_and_db}, for negative values of $\delta \beta$ with $|\delta \beta| \geq 1$ (panel (a)) the topology of the flow is similar to that of $\bv^{(2)}$, with the tilted vortex in the neighborhood of the edge of the patch, while the direction of the flow is reversed from that of $\bv^{(2)}$ (due to $\beta_w < 0$). With increasing $\delta \beta$ into positive values, the vortex disappears and at $\delta \beta \lesssim 1$ (panel (b)) the flow  resemblances $\bv^{(1)}$ (which is the total flow when $\delta \beta = 0$, just with reversed direction due to the negative $\beta_w$). However, upon further increasing $\delta \beta$, a new vortex, accompanied by a stagnant point on the $r = 0$ symmetry axis at some distance from the wall, emerges (panel (c)). This vortex is located directly above the patch, and it becomes more pronounced, and more extended into the bulk above the patch, with increasing $\delta \beta$ (accordingly, the location of the stagnant point is also displaced to larger distances from the wall), see \fref{fig:flow_v_b_and_db}(c) and (d). These results highlight the point that an alternative interpretation, which involves only induced osmotic slip, is possible for the experimental observations of tracer particles moving in opposite directions at various heights in the vicinity of the active patch. This emphasizes the necessity of additional investigations, for example comparisons with the observations in an upside down set-up, as carried out by Refs. \cite{gentile_silver-based_2020,valdez_solutal_2017}, when interpreting the drift of the tracers in terms of what is the nature of the ambient flow.

\section{Buoyancy-induced flow}
\label{sec:bulk_flow}

In view of the linearity of the incompressible Stokes equation, the flow $\bw$, which is the solution of the boundary value problem given by \eqs{eq:Stokes}, \eqref{eq:BC_u_infty}, and \eqref{eq:BC_w_wall} with a distribution of body-forces $\bff$ within the fluid, can be written in terms of the corresponding Green function  
\cite{KiKa91,Brenner_book}. For the geometry of a half-space fluid bounded by a planar wall, the Green function is known \cite{Blak71a,Blak74a}: using the notations in Ref. \cite{Spagnolie2012}, the flow $\bw_1(\bs,\bs_0)$ at the ``observation'' point $\bs$,  due to a delta force density $\bF = F \delta(\bs-\bs_0) \be_z$ at the ``source'' point $\bs_0$, is given by \cite{Spagnolie2012}
\begin{eqnarray}
\label{eq:fund_flow}
&&\bw_1(\bs,\bs_0) = \frac{F}{8 \pi \mu R} \,{\bar \bw}_1(\bar \bs,\bar \bs_0)\,,
\\
&&{\bar \bw}_1(\bar \bs,\bar \bs_0) := 
\frac{1}{|\bar \bs - \bar \bs_0|} 
\left[ \be_z + \frac{(\bar \bs - \bar \bs_0)\cdot \be_z}{|\bar \bs - \bar \bs_0|^2}
(\bar \bs - \bar \bs_0)
\right]
\nonumber\\
&& \hspace*{1.7cm} - \, \frac{1}{|\bar \bs - \bar \bs_0^*|} 
\left[ \be_z + \frac{(\bar \bs - \bar \bs_0^*)\cdot \be_z}{|\bar \bs - \bar \bs_0^*|^2}
(\bar \bs - \bar \bs_0^*)
\right]\nonumber\\
&& \hspace*{1.7cm} - 2 {\bar h}
\left[-1 + 3 \frac{((\bar \bs - \bar \bs_0^*)\cdot \be_z)^2}{|\bar \bs - \bar \bs_0^*|^2}
\right]\frac{\bar \bs - \bar \bs_0^*}{|\bar \bs - \bar \bs_0^*|^3} \nonumber\\
&& \hspace*{.5 cm} - 2 {\bar h}^2 \frac{1}{|\bar \bs - \bar \bs_0|^3} 
\left[ 
- \be_z + 
3 \frac{(\bar \bs - \bar \bs_0)\cdot \be_z}{|\bar \bs - \bar \bs_0|^2}
(\bar \bs - \bar \bs_0)
\right]\,,
\nonumber
\end{eqnarray} 
in terms of the flows corresponding to the source Stokeslet at $\bs_0$ (first line) and to the system of images located at $\bs_0^*$: a Stokeslet pointing in the opposite direction (the second line), a Stokeslet dipole of magnitude proportional to $\bar h$  (the third line), and a source dipole of magnitude proportional to ${\bar h}^2$ (the fourth line), respectively. In the above, $h := \bs_0 \cdot \be_z$, $\bs_0^* := \bs_0 - 2 h \,\be_z$, and we recall that the overbar indicates dimensionless quantities (in the case of coordinates, this is specifically by measuring in units of $R$). Since a detailed, pedagogic and illustrative discussion of the fundamental flow ${\bar \bw}_1(\bar \bs,\bar \bs_0)$ (as well as of those flows corresponding to the case of point forces oriented parallel to the wall or to the case of a free interface, rather than a wall) can be found in Ref. \cite{MEJ2018}, here we only provide an illustration of the flow ${\bar \bw}_1(\bar \bs,\bar \bs_0)$ in the \aref{sec:App_bulk_flow}. 

In terms of the fundamental flow ${\bar \bw}_1(\bar \bs,\bar \bs_0)$, the bulk driven flow due to the  distribution of body forces given by \eq{eq:f_dist} is written as
\begin{equation}
\label{eq:expr_w_flow}
\bw(\bar \bs)/W_0 =  \mathrm{sgn}(\alpha g) \int\limits_{\bar z \, > \,0} d^3 {\bar \bs}_0 \, {\bar \bw}_1(\bar \bs,{\bar \bs}_0) {\bar c}({\bar \bs}_0)\,,
\end{equation}
where the velocity scale $W_0$ of the bulk-driven flow is given by
\begin{equation}
\label{eq:def_W0}
W_0 = \frac{|\alpha g|}{8 \pi \mu} \frac{Q}{R}\,,
\end{equation}
Accordingly, the calculation of the bulk driven flow $\bw(\bs)$ reduces to the technical issue of performing a volume integral. (One notes that the axial-symmetry of the density distribution ${\bar c}({\bar \bs}_0$ with respect to the $z$-axis ensures that the flow $\bw(\bs)$ is axi-symmetric, too). Since a closed-form expression is not available for the density ${\bar c}$, at any desired ``observation'' point $\bs$ the integral above is approximated by numerical integration (using Mathematica version 13), with the function ${\bar c}({\bar r},{\bar z})$ obtained by interpolation of a uniform grid of precomputed values ${\bar c}_k$ at $100 \times 100$ points $({\bar r}_k,{\bar z}_k)$ in the relevant range $0 \leq {\bar r}, {\bar z} \leq 10$ (see \fref{fig:sol_dens_compact}), and the domain of integration restricted to the same $({\bar r},{\bar z})$ range.

The factor $\mathrm{sgn}(\alpha g)$ indicates whether the resulting body forces are pointing towards or away from the wall. Since ${\bar c}({\bar \bs}) > 0$, by \eq{eq:f_dist} the sign of $\alpha$ indicates if the inhomogeneities due to reaction lead to an increase ($\alpha > 0$) or a decrease ($\alpha < 0$) in the local mass density, respectively, while, as discussed, the sign of $g$ indicates the direction of the gravity with respect to the normal $\be_z$ of the wall. For example, for the schematic shown in \fref{fig_schematic}, for the buoyancy forces to be pointing towards the wall, as shown by the blue arrow, one either has $g < 0$ (gravity pointing towards the wall) and $\alpha > 0$, i.e., a system in which the active patch is at the bottom wall and the reaction leads to local increases in the mass density, or $g > 0$ and $\alpha < 0$, i.e., a system in which the active patch is at the top wall and the reaction leads to local decrease in the mass density. This provides the reasoning upon which experimental observations of a flow reversal upon flipping the system upside down can be --- and have been --- interpreted as evidence for buoyancy-driven flows \cite{sengupta_self-powered_2014,valdez_solutal_2017,gentile_silver-based_2020}.

\begin{figure}[!thb]
\includegraphics[width = \columnwidth]{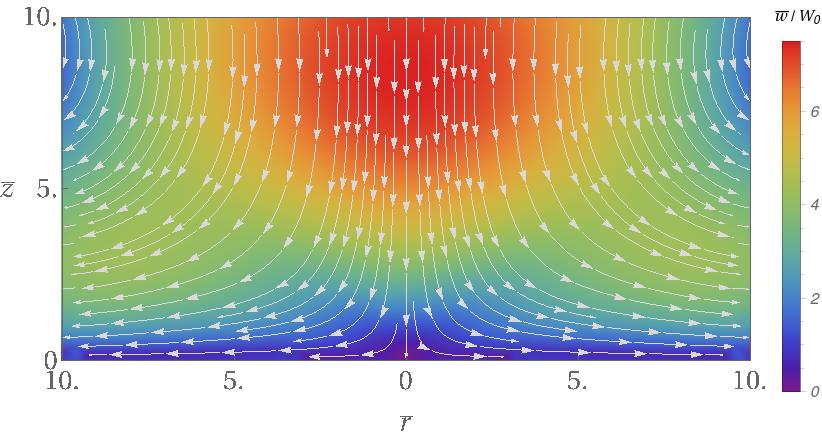}\hfill\\
\includegraphics[width = \columnwidth]{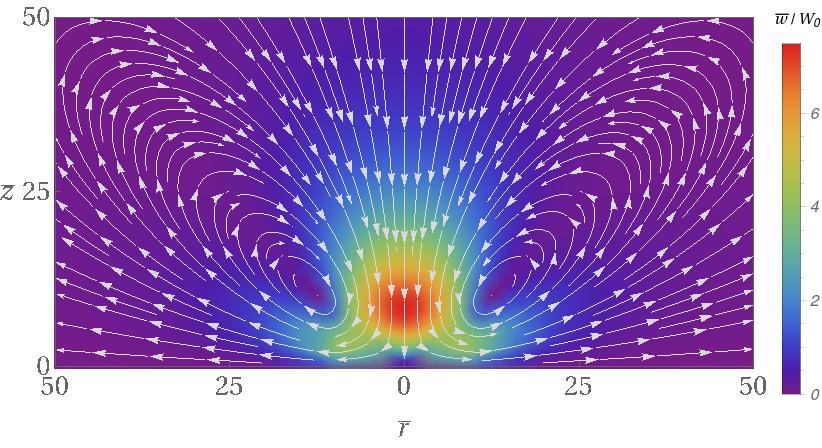}\hfill
\caption{\label{fig:flow_w} Streamlines (white) and magnitude (color coded background) for the flow ${\bar \bw}(\bar \bs)$ corresponding to the case of the compact support patch and $\mathrm{sgn}(\alpha g) = -1$ (\eqs{eq:C_gen}, \eref{eq:c_compact_patch}, \eref{eq:fund_flow}, and \eref{eq:expr_w_flow}). The top panel is a ``zoom'', from the image in the bottom panel, of the flow in a region near the active patch.  
}
\end{figure}
In \fref{fig:flow_w} we show the bulk driven flow corresponding to a situation like the one discussed above in the context of \fref{fig_schematic}, i.e., for $\mathrm{sgn}(\alpha g) = -1$ and a wall at the bottom, at length-scales of the order of the radius of the patch, i.e., $\bar r \lesssim 10, \bar z \lesssim 10$ (top panel) and also over much larger lengthscales  (bottom panel). It is clear that near the active patch, $\bar r \lesssim 2\,,~ \bar z \lesssim 2$, the bulk-driven flow has a topology similar to that of the base flow $\bv^{(1)}$ from the case of an osmotic-driven flow (compare the top panel with \fref{fig:v1_and_v2_compact} and \fref{fig:v1_and_v2_diffuse} in \aref{sec:diffuse_patch}). However, these two flows exhibit significant difference in terms of how the magnitude of the flow varies with the distance from the center of the patch, in particular near the wall, where the magnitude of the bulk-driven flow (which is anyway very small, recall that it has to vanish at the wall) decays monotonically with $r$, while the magnitude of the osmotic-driven flow exhibits a maximum at some $r \sim 1$. 

Finally, one notes that over large length-scales (bottom panel), where the spatial extent of the distributions of chemical inhomogeneities and corresponding body-forces looks ``point like'', the resulting flow  resembles the fundamental flow ${\bar \bw}_1$ (see \fref{fig:w1_flow} in \aref{sec:App_bulk_flow}). That is, as if it would be due to a ``net force'' Stokeslet located on the $z$-axis and pointing into the wall (thus the change in the direction compared to ${\bar \bw}_1$), qualitatively consistent with a far-field intuitive picture as described above.  

\section{Conclusions}
\label{sec:conc}

Motivated by the available experimental evidence that both bulk-driven and osmotic, surface-driven effects can be present within the flow induced by a realization of a chemically-active micropump, we have analyzed a simple model 
of a chemically active patch imprinted on a planar wall and analytically calculated the induced hydrodynamic flow in a Newtonian solution that occupies 
the half space above the wall. This ``semi-unbounded'' geometry corresponds to 
an experimental-cell geometry with a height much larger than its diameter, which is, in turn, much larger than the size of the patch, i.e., a weakly confined system, and complements the available numerical studies of the flow in confined geometries \cite{sengupta_self-powered_2014,ortiz-rivera_convective_2016,gao_geometric_2022,valdez_solutal_2017, afshar_farniya_sequential_2014,das_harnessing_2017,ortiz-rivera_enzyme_2016,gentile_silver-based_2020,sen_2024,esplandiu_radial_2022,niu_microfluidic_2017}. While
simplifications away from the typically complex experimental realizations --- such as  solely considering a single uncharged species and a simple ``constant flux'' activity model for the chemical reaction ---  have been necessary in order to allow analytical calculations, the resulting model permits us to obtain physically insightful results for the hydrodynamic flows induced by chemically active pumps and to understand the relative importance of each type of source of flow. We do expect that these results capture qualitative features of such chemically active systems, and thus they could provide guidance in the set-up of future experiments as well as in the interpretation of the drift-by-the-ambient-flow component of the tracer motion usually observed and studied in experimental investigations of chemically active catalytic micropumps.

For a Newtonian liquid occupying a half space above a planar wall carrying a chemically active patch with constant flux activity in an axi-symmetric geometry, we determined in closed form, as integral representations, the distribution of chemical composition inhomogeneities in the solution (\sref{sec:distri_solute_gen_sol}) as well as the Stokes flow of the solution due to the corresponding surface ``osmotic slip''  and bulk ``solutal buoyancy'' sources of flow. The general flow is written as a linear superposition of a surface-driven component $\bv$, which is generated by the base flows $\\bv^{(1)}$ and $\bv^{(2)}$ (\sref{sec:osmo_hydro}), and a bulk-driven component $\bw$ (\sref{sec:bulk_flow}), respectively. 

The base flows $\bv^{(1)}$ and $\bv^{(2)}$ are those sourced by an osmotic-slip distribution over the whole wall (including the patch region) as if the osmotic-slip coefficient would be uniform, and by a distribution with compact support at the patch and amplitude proportional to the contrast in osmotic slip coefficients between the patch and the wall, respectively. Explicit calculations of these base flows for two choices of active patches: compact sharp (\sref{sec:step_funct_activ}) or diffuse (\aref{sec:diffuse_patch}), respectively, show that the two flows do have distinct topologies, with $\bv^{(2)}$ exhibiting a recirculation vortex located near the edge of the patch. The superposition of the two flows exhibits a somewhat unexpectedly rich behavior, including qualitative changes in the topology of the flow, as a function of the contrast in chemical properties (osmotic slip coefficient) between the patch and the support wall, as shown by the example in \fref{fig:flow_v_b_and_db}. The bulk-driven flow $\bw$ exhibits a topology similar to that of $\bv^{(1)}$; however, differences between the two are clear when one considers the variations of the magnitude of the flow: $|\bv^{(1)}|$ is large near the wall, and attains a maximum value near the edge of the patch (see \fref{fig:v1_and_v2_compact}), while $|\bw|$ is maximal at some significant height (few radii of the patch) above the center of the patch and it is very small near the wall (see \fref{fig:flow_w}).  

\begin{figure*}[!thb]
\includegraphics[width = \columnwidth]{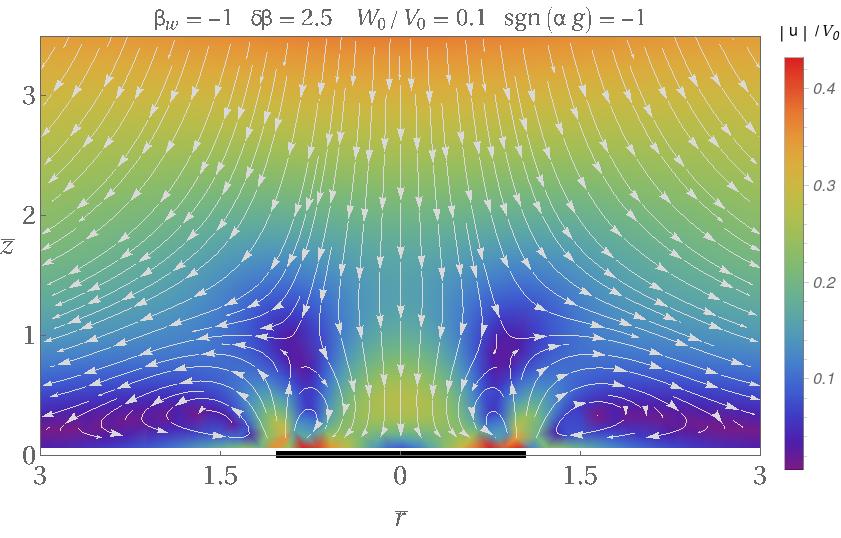}
~~
\includegraphics[width = \columnwidth]{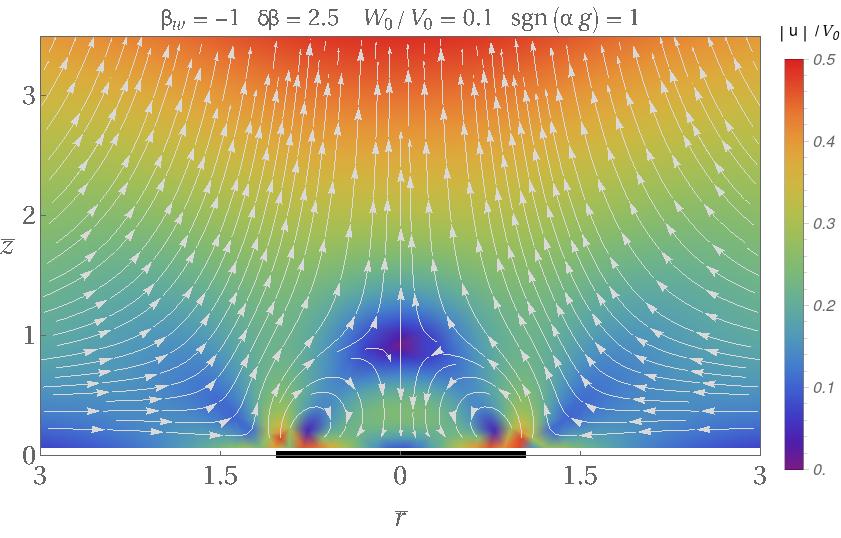}
\hfill\\
\includegraphics[width = \columnwidth]{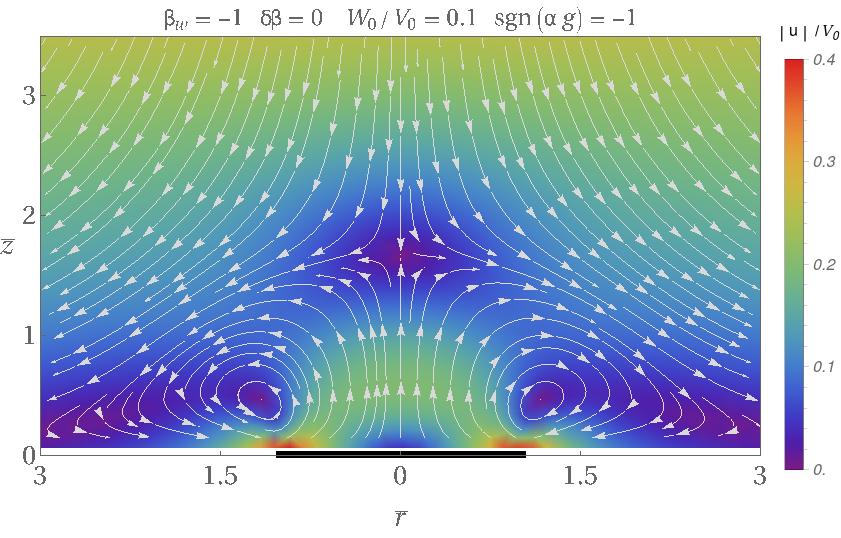}
~~
\includegraphics[width = \columnwidth]{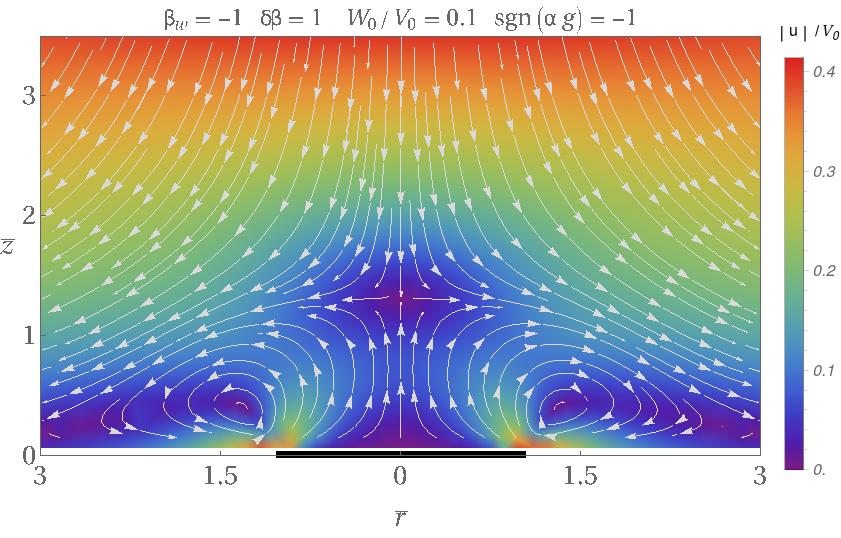}\hfill
\caption{\label{fig:total_flow_examples} Streamlines (white) and magnitude (color coded background) of the flow $\bu(\bs) = \bv(\bs) + \bw(\bs)$ corresponding to the case of the compact support patch (shown as the thick black segment on the bottom edge) resulting from various linear superpositions, as defined by the parameters above each panel, of the surface-driven flows $\bv^{(1,2)}$ and the bulk-driven flow $\bw(\bs)$.
}
\end{figure*}
In short, the surface-driven, $\bv$, and the bulk-driven, $\bw$, flows generically have different topologies and each one dominates in distinct regions 
of the fluid solution, which should facilitate experimental discrimination of these components by measurements at sufficiently distinct locations in the fluid. This is particularly useful because, as we emphasized, we expect the model and the results to capture qualitative features but, in view of the simplifications that were needed in order to obtain an analytically tractable model, attempting quantitative comparisons would be questionable even if the velocity scales could be \textit{a priori} estimated. (Such estimates are likely possible for $W_0$, as done in, e.g., Ref. \cite{valdez_solutal_2017}, but less so for $V_0$, which depends on the phenomenological osmotic-slip mobility $b$.) If at all, from the values of tracers velocity measured in experiments one can extract order of magnitude estimates for the velocity scales $W_0$ and $V_0$ interpreted as effective parameters of the model (or as functions of some effective parameters like $b$ and $\alpha$). For example, the reports of velocities $\sim~1~\mathrm{\mu m/s}$ for tracers near a wall (see, e.g., Ref. \cite{munteanu_glucose_2019}) would imply, by comparison with the scales in \fref{fig:flow_v_b_and_db} and \fref{fig:v1_and_v2_diffuse} (\aref{sec:diffuse_patch}), that $V_0 \sim 1 \,\text{--}\, 10 ~\mathrm{\mu m/s}$; similarly, measurements of tracers velocity of the order of 10~$\mathrm{\mu m/s}$ for tracers above the patch at some heights of the order of the patch radius (see, e.g., Ref. \cite{sengupta_self-powered_2014}) implies, by comparison with \fref{fig:flow_w}, an estimate $W_0 \sim \textrm{a few}~\mathrm{\mu m/s}$.

Even for this simple model of a chemically active patch and the fluid bounded just by a planar wall, in the vicinity of the patch the resulting total flow $\bu(\bs) = \bv(\bs) + \bw(\bs)$ can exhibit a rich variety of patterns in topology and magnitude when the velocity scales $V_0$ and $W_0$ are such that both $\bv(\bs)$ and $\bw(\bs)$ are relevant. We show an example of such variety of patterns in \fref{fig:total_flow_examples}, which we discuss, for clarity of argument, for the case $g<0$ (gravity pointing towards the wall). The top row illustrates how the surface-driven flow shown in \fref{fig:flow_v_b_and_db}(d) is modified by a bulk-driven flow when the reaction leads to local increases ($\alpha > 0$, top left panel) or decreases ($\alpha < 0$, top right panel) in the mass density of the solution. The bottom row illustrates that very different surface flows, with either no component ($\delta \beta = 0$) or a strong component ($\delta \beta = 1$) from $\bv^{(2)}$, respectively, can exhibit a similar topology in the presence of a bulk-driven flow; yet, they can be differentiated by the relative magnitude of the flows somewhat far above the patch and near the patch, a ratio which is much larger for the case shown in the bottom right panel.

Finally, we note that there are several directions in which the present study can be generalized. The simpler, yet very relevant experimentally, is the consideration of the same simple model considered here but in a slit-like geometry, i.e., the solution is confined by a second planar wall. This is a geometry for which the surface-driven flow can still be calculated analytically, while the bulk-driven flow at weak confinement may be well approximated by the first set of images across each wall and at strong confinement can be 
approximated by methods proposed in Ref. \cite{gentile_silver-based_2020}. 
By studying the dependence of the topology of the surface-driven and bulk buoyancy-driven flows on the height $H/R$ of the slit, such a study would connect with the investigations in Ref. \cite{Hess_2018}, which reported observations of significant qualitative changes in the flows upon increasing confinement (i.e., decreasing $H/R$). Further steps of increasing complexity could be consideration of more complex reaction schemes, e.g., a first order reaction, of multiple chemical species, and of charged species; in such cases, our expectation is that the calculation of the surface-driven flows is probably yet tractable (and that of the bulk flow not much different from above), but the calculation of the chemical composition inhomogeneities would likely be analytically untractable and would have to be carried out numerically. 

\begin{acknowledgments}
\label{Acknowledgments}
A.D.~and M.N.P.~acknowledge financial support through grants ProyExcel\_00505 funded by Junta de Andaluc{\'i}a, and PID2021-126348NB-I00 funded by MICIU/AEI/10.13039/501100011033 and ``ERDF A way of making Europe''. B.A.N., S.G., and M.N.P. acknowledge financial support from the Executive Agency for Higher Education, Research, Development and Innovation Funding of Romania grant number PN-III-P4-PCE-2021-1231 (contract no. PCE 21 from 2022). M.N.P.~also acknowledges support from
Ministerio de Universidades de Espa{\~n}a through a Mar{\'i}a Zambrano grant.
W. E. U. gratefully acknowledges the technical support and advanced computing resources from University of Hawaii Information Technology Services – Cyberinfrastructure, funded in part by the National Science Foundation CC* awards \#2201428 and \#2232862, the partial funding from the Donors of the American Chemical Society Petroleum Research Fund, grant number 60809-DNI9, and the partial funding from the Army Research Office under Grant Number W911NF-23-1-0190. The views and conclusions contained in this document are those of the authors and should not be interpreted as representing the official policies, either expressed or implied, of the Army Research Office or the U.S. Government.
\end{acknowledgments}



\appendix
\counterwithin{figure}{section}
\setcounter{figure}{0}

\section{Experiment: materials and methods}
\label{sec:mater_meth}

\noindent\textit{Materials}: GOX (Cat. No. G7141), glucose (Cat. No. G7021), bovine serum albumin (BSA, Cat. No. G4287), 3-aminopropyltri-ethoxysilane (APTES, Cat. No. A3648), glutaraldehyde (25\%, Cat. No. G400-4), silica microparticles (3 $\mathrm{\mu m}$ in diameter, Cat. No. 66373), and glycerol (Cat. No. G7893) were purchased from Sigma Aldrich Inc. Acetone (min. 99.92\%) and ethanol (min 99.5\%) were purchased from the local distributor Chimreactiv SRL (Bucharest, Romania). Toluene (Cat. No. 5191), high-precision microscope cover glasses (24 mm $\times$ 60 mm $\times$ 0.17 mm, Cat. No. 277684063, and 22 mm $\times$ 22 mm $\times$ 0.17 mm, Cat. No. 41022012) and polystyrene Petri dishes (3.5 cm diameter, Cat. No. EL46.1) were purchased from Carl Roth GmbH. Borosilicate glass capillaries (100 mm $\times$ 1.5 mm $\times$ 0.86 mm, Cat. No. GC150F-10) were purchased from Harvard Apparatus. All (bio)chemicals have been used as received. All necessary solutions have been prepared with ultrapure water from a Millipore Direct-Q3 UV water purification system.

\noindent\textit{Fabrication of the GOX patches}: GOX patches were made by slightly modifying a previously reported procedure \cite{munteanu_glucose_2019,munteanu_impact_2021}. First, the planar walls, on which the GOX patches will be imprinted, are prepared. The polystyrene wall is obtained by simply oxidizing a commercially available 3.5 cm diameter polystyrene Petri dish in a PSD-UV surface decontamination system (from Novascan, USA) for 15 min. To obtain the glass-based wall, a microscope cover glass (24 mm × 60 mm × 0.17 mm) is washed with acetone and ethanol and subsequently modified with APTES by immersion into an APTES solution (5\%, made in toluene) for 10 min. The APTES-modified cover glass is further modified with BSA by incubating it with a BSA solution (3.3 mg/mL) for 15 min. After washing the BSA-modified cover glass with water, the BSA coating is stabilized via cross-linking through exposure for 15 min to the vapors of a 25\% aqueous solution of glutaraldehyde. Next, a GOX-based ``ink'' is prepared by mixing 2 parts of a GOX solution (50 mg/mL, prepared in an aqueous solution with 1\% glycerol) with 1 part of a BSA solution (50 mg/mL, prepared in an aqueous solution with 1\% glycerol). A glass micropipette, made out of a borosilicate glass capillary, is then loaded by capillarity with a few microliters of the GOX ink and subsequently brought into contact with the planar wall (i.e., with either the UV-oxidized polystyrene surface or the BSA-modified glass surface). This way, a small volume of the GOX ink is transferred from the pipette onto the planar wall; the deposited GOX ink dries within one-two seconds. The enzyme patch, imprinted by using the procedure describe above, is then stabilized via cross-linking through exposure for 15 min to the vapors of a 25\% aqueous solution of glutaraldehyde. Finally, before the start of an experiment, the micropump is thoroughly washed with water to remove any components which were not strongly anchored to the UV-oxidized polystyrene surface or the BSA-modified glass surface. Some of the GOX patches made onto the UV-oxidized surfaces were also coated with BSA. This was done by a second step of incubating the surface (on which the GOX patch is already imprinted) with a BSA solution (3.3 mg/mL) for 15 min.

The three GOX patches made onto BSA-coated glass surfaces and used in the present study were characterized by an average diameter of 867 $\pm$ 29 $\mathrm{\mu m}$. The three GOX patches made onto UV-oxidized polystyrene surfaces and used in the present study were characterized by an average diameter of 1130 $\pm$ 56 $\mathrm{\mu m}$. 

The way tracer particles behave, in the close proximity of the enzyme patches, was documented via optical microscopy through the bottom wall of the cylindrical measuring cell (1~cm diameter and 1~mm height) by using an Observer D1 inverted microscope from Carl Zeiss AG equipped with an EC-Epiplan, 5$\times$/0.13 HD objective from the same company and with a DFK 33UX264 CCD camera from The Imaging Source Europe GmbH. Video recordings of the motion of the tracer particles, when the patches were active, have been carried out at 2 frames per second using IC Capture (from the producer of the mentioned CCD camera) as image acquisition software.

\setcounter{figure}{0}

\section{Example: the case of a diffuse patch.}
\label{sec:diffuse_patch}

In order to verify whether or not the results depend qualitatively on the exact shape of the patch, or on the compact-support vs diffuse nature of the distribution of the activity (see the inset in \fref{fig_schematic}), we have calculated the hydrodynamic flow due to the activity induced osmotic slip for the case of the activity distribution 
\begin{equation}
 \label{eq:diffuse_patch}
\mathcal{I}(\bar r) = e^{-{\bar r}^2}\,,
\end{equation}
which does not exhibit sharp edges but vanishes exponentially fast, with a characteristic lengthscale $R$, with the distance $r$ from the center of the  patch; for this choice, the surface integrated activity, i.e., the monopole of the source of solute, is the same as that of the step-function activity discussed in \sref{sec:step_funct_activ}. The steps in the calculations are the same as in \sref{sec:step_funct_activ}; consequently, here we just summarize the results.
 
\begin{figure}[!b]
\includegraphics[width = .99\columnwidth]{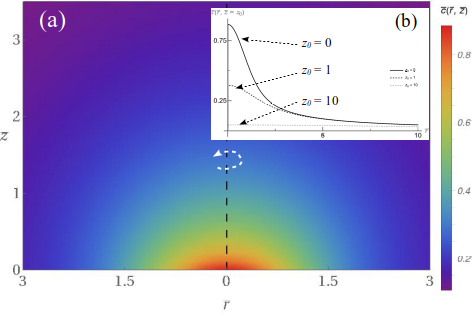}
\caption{\label{fig:sol_dens_diffuse} (a) Color coded (bar scale at the right) dimensionless density of solute ${\bar c}(\bar r, \bar z)$, \eqs{eq:C_gen} and \eref{eq:c_diffuse_patch}, for the case of a diffuse patch (\eq{eq:diffuse_patch}); the curved white arrow is a reminder of the axial symmetry around O$z$. (b) In plane dimensionless density of solute ${\bar c}(\bar r,\bar z =z_0)$ at various heights $z_0$ above the wall.
}
\end{figure}
For this activity function, \eq{eq:coef_c_gen} renders 
\begin{equation}
\label{eq:c_diffuse_patch}
\mathfrak{c}(\xi) = \frac{1}{2} e^{-\xi^2/4}
\end{equation}
for the coefficients in the integral representation of the density. The integral representation cannot be further simplified; accordingly the density $\bar c(\bar r,\bar z)$ is calculated by performing the integral numerically, using the software Mathematica (version 13) at any given point $\bar \bs = (\bar r,\bar z)$. The result, shown in \fref{fig:sol_dens_diffuse}, 
exhibits the expected $1/|\bs|$ asymptotic decay with the distance from the patch (corresponding to the far field interpretation as the diffusion field from a monopolar  source at the origin).

\begin{figure}[!h]
\includegraphics[width=0.95\columnwidth]{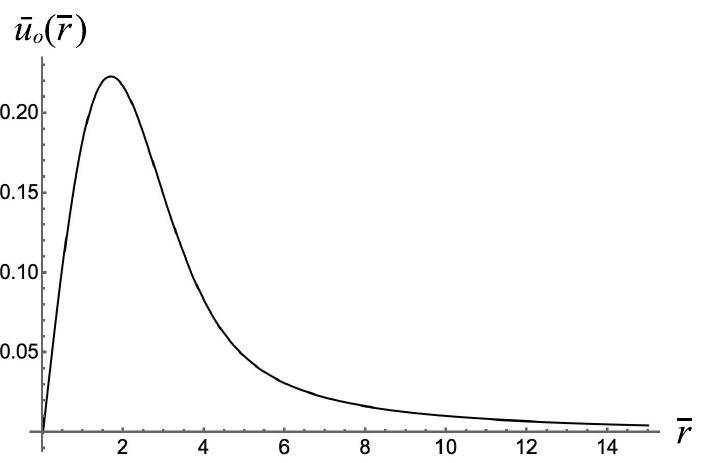}
\includegraphics[width=0.95\columnwidth]{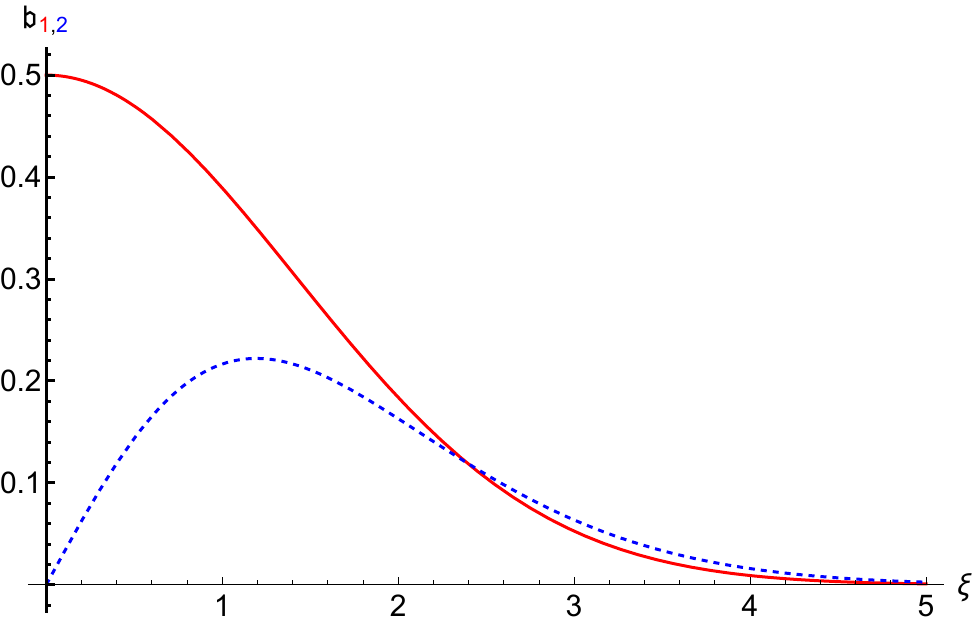}
\caption{\label{fig:slip_and_Bs_diffuse_patch} The dimensionless osmotic slip ${\bar u}_0(\bar r)$ (top panel) and the coefficients $\mathfrak{b}_{1,2}(\xi)$ (bottom panel) for the diffuse patch distribution in \eq{eq:diffuse_patch}. 
}
\end{figure}

The corresponding induced active osmotic slip, \eqs{eq:slip_vel_V0}, which is calculated by using the radial derivative of ${\bar c}(\bar r, \bar z = 0)$ because there are no mathematical issues with the evaluation at the wall, is shown in the top panel of \fref{fig:slip_and_Bs_diffuse_patch}. Qualitatively, it is similar with that occurring in the case of the step-function activity: it has a pronounced peak at moderate $\bar r$, and vanishes fast with the distance from the center of the patch.

Turning to the basic flows $\bv^{(1,2)}$, these are calculated numerically with a similar procedure as in \sref{sec:step_funct_activ}: $\bv^{(1)}$ directly, exploiting that $\mathfrak{b}_1(\xi) = \mathfrak{c}(\xi)$ is known in closed form; $\bv^{(2)}$ by using an upper cut-off $\xi_M = 10$ (which is sufficient, due to the exponential decay of $\mathfrak{b}_2(\xi)$, see \fref{fig:slip_and_Bs_diffuse_patch}), then determine $\mathfrak{b}_2(\xi)$, with $0\leq \xi \leq \xi_M$, by interpolation of a list of $250$ pre-calculated values $\mathfrak{b}_2(\xi_k)$, with $\xi_k$ uniformly spaced in the interval $0 \leq \xi \leq \xi_M$, and plugging the resulting interpolating function $\mathfrak{b}_2(\xi)$ into the numerical computation of the corresponding integrals involved in the $r$ and $z$ components of $\bv_2$. 

\begin{figure}[!thb]
\includegraphics[width = \columnwidth]{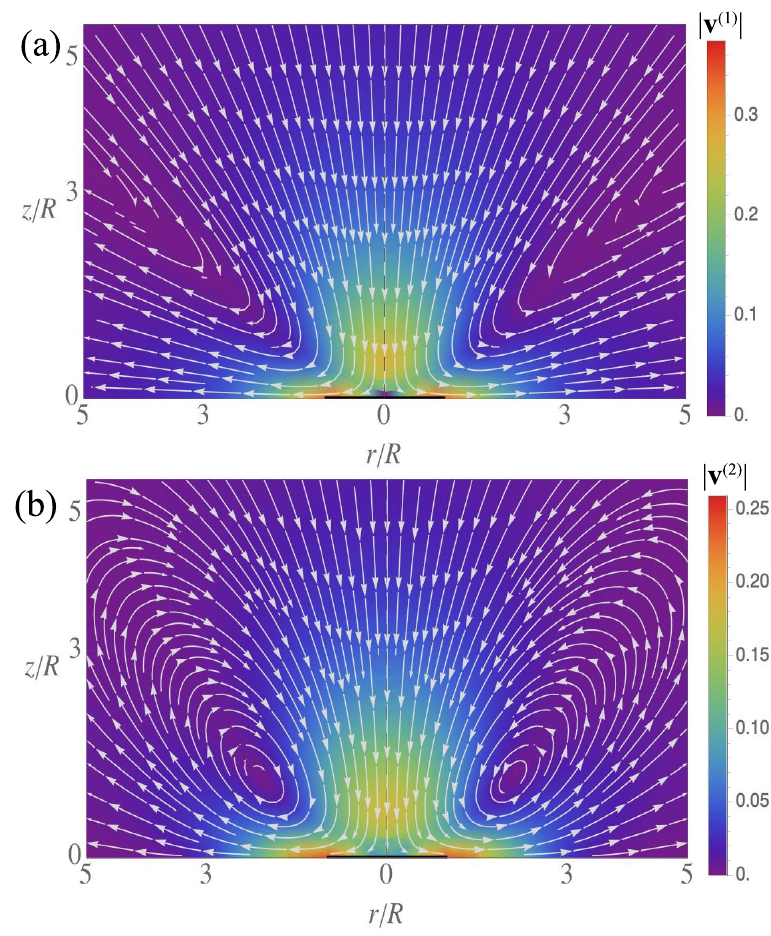}
\caption{\label{fig:v1_and_v2_diffuse} Streamlines (white) and magnitude (color coded background) for the basic flows $\bv^{(1)}(\bar \bs)$ (panel (a)) and $\bv^{(2)}(\bar \bs)$ (panel (b)) driven by the osmotic slip induced by a diffuse  active patch (\eqs{eq:v1_v2}, \eref{eq:b_func}, and \eref{eq:c_diffuse_patch}).    
}
\end{figure}
The result of such a computation on a rectangular grid in the range $0\leq \bar r \leq 5$ and $5 \times 10^{-2} \leq \bar z \leq 5$ is shown in \fref{fig:v1_and_v2_diffuse}(a) as white streamlines on a color background that encodes the magnitude 
$|{\bv}_{1,2}|$, respectively, of the corresponding flow.  As anticipated in the brief discussion in \sref{sec:model}, the flows $\bv_{1,2}$ due to the activity of a diffuse patch, \eq{eq:diffuse_patch}, are qualitatively similar to those occurring in the case of a step-function distribution of activity discussed in the main text. Accordingly, the corresponding discussion and conclusions from \sref{sec:step_funct_activ} regarding the behavior of $\bv/V_0 = \beta_w \bv_{1} + \delta \beta \bv_{2}$ directly translates for the diffuse patch.


\setcounter{figure}{0}

\section{Illustrations of the fundamental solution ${\bar w}_1$ for buoyancy (body force distributions) flows}
\label{sec:App_bulk_flow}

\begin{figure}[!htb]
\includegraphics[width = \columnwidth]{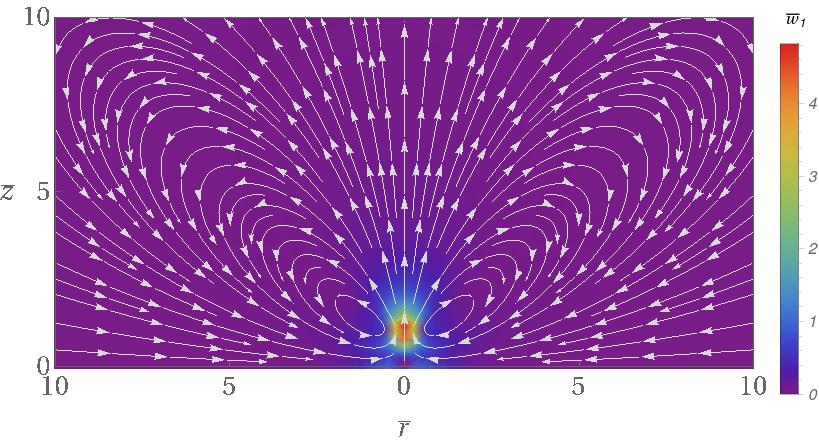}
\caption{\label{fig:w1_flow} Streamlines (white) and magnitude (color coded background) of the fundamental flow ${\bar \bw}_1(\bar \bs,{\bar \bs}_0)$ induced by a point force, oriented in the positive $z-$direction, on the $z$-axis at ${\bar \bs}_0 = (\bar r_0 = 0, \bar z_0 = 1)$.    
}
\end{figure}

\Fref{fig:w1_flow} illustrates the fundamental flow  ${\bar \bw}_1$ for a Stokeslet, pointing in the positive $z$-direction, located on the  $z$-axis at $\bar z_0 = 1$.

\bibliography{refs_Intro,refs_theory}

\clearpage

 \section*{Supplementary Material}
The Supplementary Material consists of four video recordings. (They can be obtained by contacting the author SG.) These recordings show patches made of glucose oxidase (GOX) and the motion of tracer particles carried by the surface-driven (osmotic) flows induced by such patches in different experimental conditions. The video recordings are described in Table S1 below.
\begin{figure}[!th]
\includegraphics[width = \columnwidth]{Supl_Mater_table}
\end{figure}
\clearpage

\end{document}